\documentclass[structabstract]{aa}  
\usepackage{graphicx}
\usepackage{txfonts}
\begin{document}
\title{Chemical surface inhomogeneities in late B-type stars with Hg and Mn 
peculiarity}

\subtitle{I. Spot evolution in HD 11753 on short and long time 
scales\thanks{Based on observations obtained with the CORALIE {\'E}chelle 
Spectrograph on the 1.2-m Euler Swiss telescope, situated at La Silla, Chile; 
based on observations made with ESO Telescopes at the La Silla Paranal 
Observatory under programmes 076.D-0172 and 077.D-0477; and based on data 
obtained from the ESO Science Archive Facility under request number
HHKorhonen15448}$^,$\thanks{Table 2 is only available in electronic form
at the CDS via anonymous ftp to cdsarc.u-strasbg.fr (130.79.128.5)
or via http://cdsweb.u-strasbg.fr/cgi-bin/qcat?J/A+A/}}

\author{H.\,Korhonen\inst{1,2,3}
  \and
  J.F.\,Gonz\'alez\inst{4}
  \and
  M.\,Briquet\thanks{F.R.S.-FNRS Postdoctoral Researcher, Belgium}\inst{5,6}
  \and
  M.\,Flores Soriano\inst{7}
  \and
  S.\,Hubrig\inst{7}
  \and
  I.\,Savanov\inst{8}
  \and
  T.\,Hackman\inst{3,9}
  \and
  I.V.\,Ilyin\inst{7}
  \and
  E.\,Eulaers\inst{5}
  \and
  W.\,Pessemier\inst{6}
}

\institute{
  Niels Bohr Institute, University of Copenhagen, Juliane Maries Vej 30, 
  DK-2100 Copenhagen, Denmark\\
  \email{heidi.korhonen@nbi.ku.dk} 
  \and
  Centre for Star and Planet Formation, Natural History Museum of Denmark, 
  University of Copenhagen, {\O}ster Voldgade 5-7, DK-1350 Copenhagen, Denmark
  \and
  Finnish Centre for Astronomy with ESO (FINCA), University of 
  Turku, V{\"a}is{\"a}l{\"a}ntie 20, FI-21500 Piikki{\"o}, Finland
  \and
  Instituto de Ciencias Astronomicas, de la Tierra, y del Espacio (ICATE), 
  5400 San Juan, Argentina
  \and
  Institut d'Astrophysique et de G{\'e}ophysique, Universit{\'e} de 
  Li{\`e}ge, All{\'e}e du 6 ao{\^u}t 17, Sart-Tilman, B{\^a}t. B5C, 4000 
  Li{\`e}ge, Belgium 
  \and
  Instituut voor Sterrenkunde, Katholieke Universiteit Leuven, 
  Celestijnenlaan 200 D, B-3001 Leuven, Belgium
  \and 
  Leibniz-Institut f{\"u}r Astrophysik Potsdam (AIP), An der Sternwarte 16, 
  14482 Potsdam, Germany
  \and
  Institute of Astronomy, Russian Academy of Sciences, Pyatnitskaya 48, 
  Moscow 119017, Russia
  \and
  Department of Physics, PO Box 64, FI-00014 University of Helsinki, Finland
}

\date{Received September 15, 1996; accepted March 16, 1997}

% \abstract{}{}{}{}{} 
% 5 {} token are mandatory

\abstract
    {}
    {Time series of high-resolution spectra of the late B-type star HD~11753 
      exhibiting HgMn chemical peculiarity are used to study the surface 
      distribution of different chemical elements and their temporal evolution.}
    {High-resolution and high signal-to-noise ratio spectra were obtained 
      using the CORALIE spectrograph at La Silla in 2000, 2009, and 2010.  
      Surface maps of \ion{Y}{ii}, \ion{Sr}{ii}, \ion{Ti}{ii}, and \ion{Cr}{ii} 
      were calculated using the Doppler imaging technique. The results were 
      also compared to equivalent width measurements. The evolution of chemical 
      spots both on short and long time scales were investigated.}
    {We determine the binary orbit of HD~11753 and fine-tune the rotation
      period of the primary. The earlier discovered fast evolution of the
      chemical spots is confirmed by an analysis using both the chemical spot 
      maps and equivalent width measurements. In addition, a long-term decrease
      in the overall \ion{Y}{ii} and \ion{Sr}{ii} abundances is discovered. A 
      detailed analysis of the chemical spot configurations reveals some 
      possible evidence that a very weak differential rotation is operating in 
      HD~11753.}
    {}

\maketitle
%
%________________________________________________________________

\section{Introduction}

A fraction of late B-type stars, the so-called HgMn stars, exhibit enhanced 
absorption lines of certain chemical elements, notably Hg and Mn, and 
underabundance of He. About 150 stars with the HgMn peculiarity are currently 
known (Renson \& Manfroid \cite{ren_man09}), and the elements with anomalously 
high abundances in HgMn stars are known to be distributed inhomogeneously over 
the stellar surface (e.g., Adelman et al. \cite{adelman}). Still, in contrast 
to magnetic chemically peculiar stars with predominantly bipolar magnetic field 
structure, no strong large-scale, organised magnetic field of kG order has ever 
been detected in HgMn stars. Therefore, also the mechanisms responsible for the 
development of the chemical anomalies in HgMn stars are poorly understood. 

Using the Doppler Imaging technique, Kochukhov et al.\ (\cite{kochukhov07}) 
report a discovery of the secular evolution of the mercury distribution on the 
surface of the HgMn SB2 star $\alpha$~And. Until very recently, the only other 
HgMn star with a published surface elemental distribution has been AR\,Aur 
(Hubrig et al. \cite{hubrig06_ARAur}; Savanov et al. \cite{savanov09}; Hubrig 
et al. \cite{hubrig10}), where the discovered surface chemical inhomogeneities 
are related to the relative position of the companion star. A slow evolution of
the chemical spots is also seen in AR\,Aur (Hubrig et al. \cite{hubrig10}), and
for the first time, a fast dynamical evolution of chemical spots on a time 
scale of a few months has been reported on the HgMn-type binary HD\,11753 by 
Briquet et al. (\cite{briquet}, from hereon Paper\,1). This evolution implies 
hitherto unknown physical mechanisms operating in the outer envelopes of late 
B-type stars with HgMn peculiarity. Doppler maps of HD~11753 for \ion{Y}{ii}, 
\ion{Sr}{ii}, \ion{Ti}{ii}, and \ion{Cr}{ii} obtained using HARPSpol data have 
been recently presented by Makaganiuk et al. (\cite{maka12}) for one epoch.

Magnetic fields up to a few hundred Gauss have been detected in several
HgMn stars using FORS\,1 low-resolution spectropolarimetry (Hubrig et al.
\cite{hubrig06_AN}). These detections were not confirmed by Bagnulo et al. 
(\cite{bagnulo}). Furthermore, measurements of the magnetic field with the 
moment technique (e.g., Mathys \cite{mathys91}, \cite{mathys95a}, 
\cite{mathys95b}) using several elements in a circularly polarised 
high-resolution spectrum of AR\,Aur revealed a longitudinal magnetic field of 
the order of a few hundred Gauss in both stellar components and a quadratic 
field of the order of 8\,kG on the surface of the primary star (Hubrig et al. 
\cite{hubrig10}). On the other hand, the investigation of high-resolution,
circularly polarised spectra using the least-squares deconvolution (LSD) 
technique by Makaganiuk et al. (\cite{maka11}) did not reveal any signatures 
of global magnetic fields in 41 HgMn stars, anymore than did the LSD 
investigations of AR\,Aur by Folsom et al. (\cite{folsom}) and of HD~11753 by 
Makaganiuk et al. (\cite{maka12}). Almost all studies finding no magnetic 
fields in HgMn stars used in the LSD technique all the lines from all the 
elements simultaneously in the analysis. In a very recent study by Hubrig et 
al. (\cite{Hubrig_mg}), the HARPSpol data used by Makaganiuk et al. 
(\cite{maka11}) have been reanalysed. The magnetic field was measured with the 
moment technique using spectral lines of several elements separately, achieving
detections of magnetic fields of up to 60--80\,G in the highest signal-to-noise
ratio (S/N) spectra of a few HgMn stars (Hubrig et al. \cite{Hubrig_mg}). 
Additional data and analysis are needed to definitely draw a conclusion about 
the presence of a magnetic field in these objects.

Also vertical abundance anomalies have been reported in the atmospheres of 
HgMn stars. Savanov \& Hubrig (\cite{sav_hub}) report an increase in the Cr
abundance with height in the stellar atmosphere in nine HgMn stars. In 
addition, they report indications of strongest vertical gradients occurring in
the hotter stars. Recently, Thiam et al. (\cite{thiam}) have investigated 
possible vertical stratification in four HgMn stars using five chemical 
elements. For most stars and elements, no evidence of radial abundance 
anomalies has been found, except for one case, HD~178065, where the Mn 
abundance shows clear evidence of increasing abundance with depth.

In this first paper in a series of papers on HgMn stars, we continue the
investigation of HD~11753 started in Paper\,1. Here, CORALIE data from 2009 and
2010 and HARPSpol data from 2010 are used in the analysis. The series will be
continued with studies of other HgMn stars for which time series of spectra are
already at our disposal. The target of the current paper, HD~11753, is a 
single-lined spectroscopic binary with an effective temperature of 10612~K 
(Dolk et al.\ \cite{dolk03}). Here, the binary orbit is determined using 
all the available data, and a new period determination is carried out using all
the data from 2000 (Paper\,1), 2009, and 2010, and a few datapoints 
from other epochs. Chemical surface distribution maps using the fine-tuned 
period are presented for \ion{Y}{ii}, \ion{Ti}{ii}, \ion{Sr}{ii}, and 
\ion{Cr}{ii} for four epochs. Tests are carried out to show the 
reliability of the results on the evolution of chemical spots on the surface of
this star.

\section{Observations}

The high-resolution spectra of HD\,11753 used in this work were obtained using 
the CORALIE {\'e}chelle spectrograph at the 1.2m Leonard Euler telescope in La
Silla, Chile. The data from 2000 were already used in Paper\,1. The new data for
the year 2009 consist of 23 spectra obtained from July 30 to August 13, 2009 
and the 2010 data of 17 spectra obtained from January 16 to 30, 2010. 

\onltab{1}{
\begin{table}
\caption{CORALIE observations of HD\,11753 in 2009 and 2010.}
\label{obs}
\centering 
\begin{tabular}{c c c r}
\hline\hline            
Obs Date & HJD          & Rotational     & S/N \\ 
         & 2\,455\,800+ & Phase$^{\rm a)}$  &  at $\lambda\lambda$\,4200~{\AA}\\
\hline    
\multicolumn{4}{c}{August 2009, CORALIE} \\
2009-07-31 & 43.8735 & 0.357 & 215 \\
2009-07-31 & 43.9161 & 0.361 &  96 \\
2009-08-01 & 44.8585 & 0.460 & 272 \\
2009-08-02 & 45.8368 & 0.563 & 189 \\
2009-08-02 & 45.9191 & 0.572 & 145 \\
2009-08-03 & 46.7864 & 0.663 & 209 \\
2009-08-03 & 46.8358 & 0.668 & 113 \\
2009-08-03 & 46.8807 & 0.672 & 105 \\
2009-08-04 & 47.7124 & 0.760 & 138 \\
2009-08-04 & 47.8853 & 0.778 & 172 \\
2009-08-05 & 48.7595 & 0.870 & 133 \\
2009-08-05 & 48.9110 & 0.885 & 171 \\
2009-08-06 & 49.7321 & 0.972 & 132 \\
2009-08-06 & 49.9238 & 0.992 & 195 \\
2009-08-09 & 52.7933 & 0.293 & 218 \\
2009-08-09 & 52.9318 & 0.307 & 368 \\
2009-08-10 & 53.7751 & 0.396 & 174 \\
2009-08-10 & 53.8770 & 0.406 & 229 \\
2009-08-11 & 54.7410 & 0.497 &  42 \\
2009-08-11 & 54.9260 & 0.517 & 179 \\
2009-08-12 & 55.9140 & 0.620 & 197 \\
2009-08-13 & 56.7986 & 0.713 & 138 \\
2009-08-13 & 56.8963 & 0.723 & 236 \\
\multicolumn{4}{c}{January 2010, CORALIE} \\
2010-01-16 & 212.5491 & 0.055 & 184 \\
2010-01-16 & 212.6276 & 0.063 & 158 \\
2010-01-17 & 213.5329 & 0.158 & 386 \\
2010-01-17 & 213.6196 & 0.167 & 202 \\
2010-01-18 & 214.5356 & 0.263 & 135 \\
2010-01-18 & 214.6220 & 0.272 & 277 \\
2010-01-19 & 215.5371 & 0.368 & 172 \\
2010-01-19 & 215.6143 & 0.376 & 212 \\
2010-01-20 & 216.5348 & 0.473 & 236 \\
2010-01-21 & 217.5339 & 0.578 & 206 \\
2010-01-23 & 219.5366 & 0.788 & 280 \\
2010-01-24 & 220.5373 & 0.893 &  79 \\
2010-01-25 & 221.5935 & 0.004 & 151 \\
2010-01-27 & 223.5975 & 0.214 & 351 \\
2010-01-28 & 224.5336 & 0.312 & 315 \\
2010-01-29 & 225.5659 & 0.421 & 292 \\
2010-01-30 & 226.5320 & 0.522 & 197 \\
\hline                       
\end{tabular}
\tablefoot{a) calculated using P=9.53077 and T=2\,451\,800.0}
\end{table}
}

\begin{figure}
\centering
\includegraphics[width=0.23\textwidth]{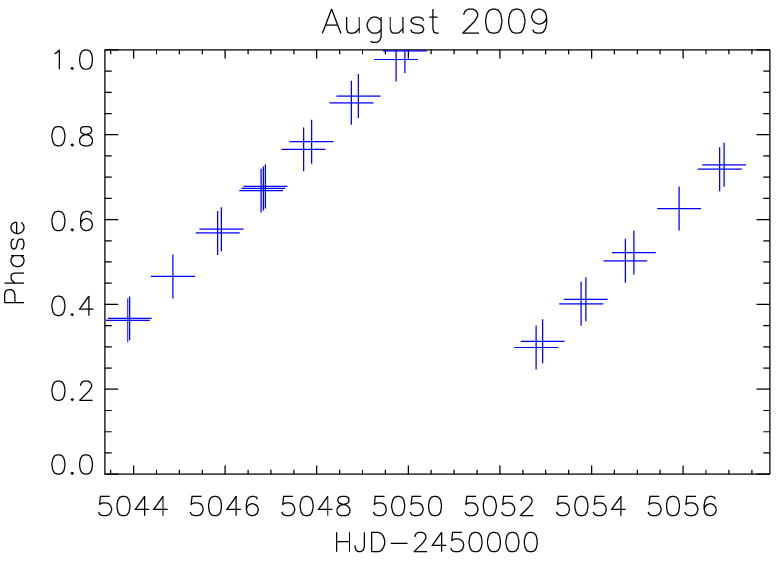}
\includegraphics[width=0.23\textwidth]{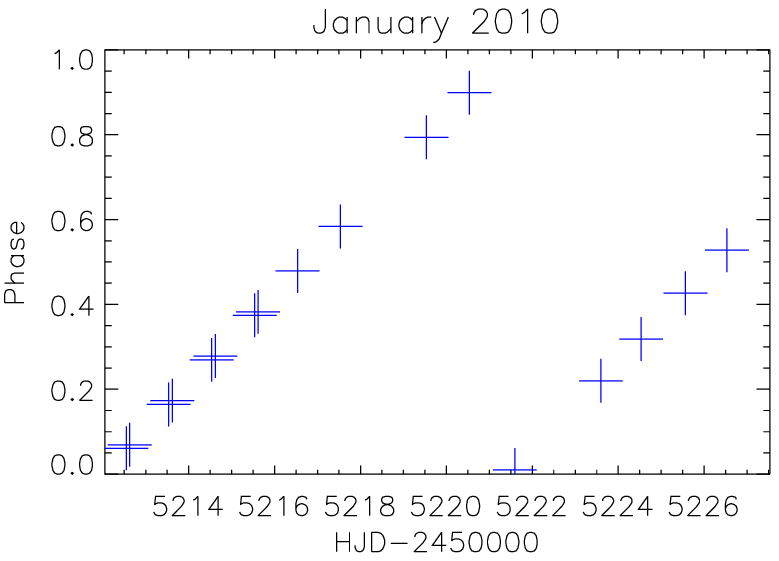}
\caption{Phase coverage for the observations in August 2009 (left) and 
January 2010 (right) CORALIE observations.}
\label{phases}
\end{figure}

The wavelength coverage of the CORALIE spectrograph is from 3810\,{\AA} to 
6810\,{\AA}, recorded in 68 orders. The CCD camera is a 2k x 2k device with 
pixels of 15$\mu$m. In July 2007 an upgrade of the instrument was carried out, 
increasing the throughput by factors 6 to 8 and spectral resolution by 10\% to
20\% (see S{\'e}gransan et al. \cite{coralie}). This results in a spectral 
resolution of approximately 55,000, and the mean S/N at wavelength 4200~{\AA} 
of 259 in the 2009 dataset and of 214 in the 2010 observations. More details on 
the observations, including the measured S/Ns, are given in the on-line 
Table\,\ref{obs} and the phases are also shown in Fig.\,\ref{phases}. 

For the data reduction the CORALIE on-line reduction package with the usual 
steps of bias removal, flat-fielding, background subtraction, and wavelength 
calibration using ThAr calibration lamp is used. The final normalisation 
was done to the pipeline-merged spectrum using a cubic spline fit.

In the period analysis we also include five spectra taken in October 2005 and 
August 2006 with the FEROS {\'e}chelle spectrograph on the 2.2m telescope at 
the ESO La Silla observatory. These spectra cover the range 3530--9220 \AA \ 
with a nominal resolving power of 48\,000.

In addition, we also use the HARPSpol observations of HD11753, which are 
publicly available in the ESO archive. HARPSpol is a polarimeter (Snik et al. 
\cite{HARPSpol}) feeding the HARPS spectrometer (Mayor et al. \cite{HARPS}) at 
the ESO 3.6m telescope in La Silla. More information on the observations 
themselves can be obtained in Makaganiuk et al. (\cite{maka12}), and on the 
data reduction in Hubrig et al. (\cite{Hubrig_mg}).

\section{Orbital analysis}

HD\,11753 has been known to be radial velocity variable since the first 
spectrographic observations taken one century ago by Moore (\cite{m11}) and 
then by Campbell \& Moore (\cite{lick}). Its variability was confirmed by 
Dworetsky et al. (\cite{d82}), who estimated the amplitude to be at least 
12--15 km/s, but they were unable to determine the period. Combining these 
observations with new measurements, Leone \& Catanzaro (\cite{lc99}) calculated
a spectroscopic orbit with a period of 41.49 days and a semi-amplitude of 9 
km/s.

The long series of high-quality observations taken in the last few years are 
not consistent with the published orbit. We therefore carried out a general 
analysis of all the available observations and found that the orbit in fact has
a period of 1126 days. Radial velocities for the 153 CORALIE spectra, five 
FEROS spectra, and 13 HARPS spectra were measured by cross-correlations using 
a synthetic spectrum for $T_{\rm eff}$=11000~K and $\log g$=4.0 as template. We 
also measured two mid-resolution (R=13000) spectra obtained with the REOSC 
spectrograph at the 2m telescope of the CASLEO (Complejo Astron\'omico El 
Leoncito, San Juan, Argentina).

Table 2, available at the CDS, lists our 173 radial velocity measurements and 
contains the following information: Heliocentric Julian Date, orbital phase, 
radial velocity measurement, error of the measurement, residual 
observed-minus-calculated and the instrument information (H=HARPS, C=CORALIE, 
F=FEROS, R=REOSC). In addition we included three datasets from the literature 
in our analysis: six measurements from Leone  \& Catanzaro (\cite{lc99}), nine 
high-resolution spectrographic observations taken by Dworetsky et al. 
(\cite{d82}) with the coud\'e spectrograph of the Palomar 5m telescope, and 11 
observations of lower resolution taken by the same authors with the coud\'e 
spectrograph of the 1.9m telescope at SAAO.

\begin{figure}
\centering
\includegraphics[width=0.42\textwidth,height=0.5\textwidth]{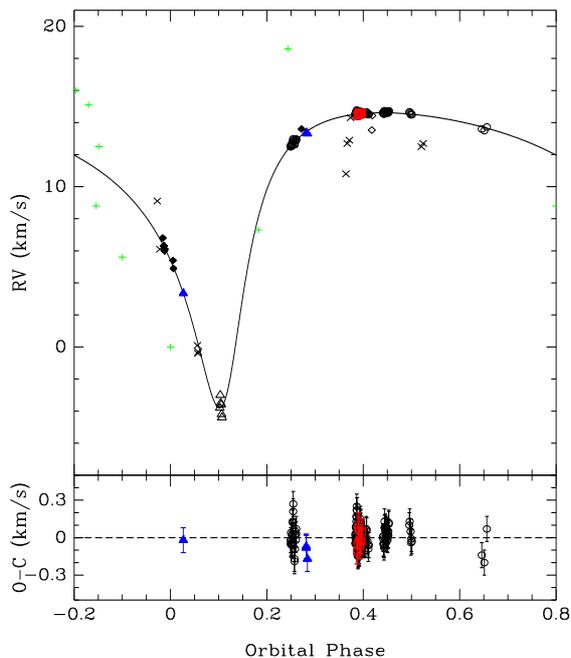}
\caption{Upper panel of the plot shows the radial velocity curve. Symbols are 
as follows: circles = CORALIE, (red) squares = HARPS, (blue) filled triangles =
FEROS, open triangles = Leone \& Catanzaro (\cite{lc99}), filled diamonds = 
Palomar plates of Dworetsky et al. (\cite{d82}), crosses = SAAO plates of 
Dworetsky et al. (\cite{d82}), and (green) pluses = Campbell \& Moore 
(\cite{lick}). Residuals observed-minus-calculated for our high-resolution 
measurements are shown in the lower panel.}
\label{fig:rvcurve}
\end{figure}

The Lomb-Scargle method (Press \& Rybicki \cite{pdg}) was applied to identify 
the period and a Keplerian orbit was fitted by the least squares method, 
obtaining the parameters listed in Table\,\ref{tab:orbit}. The radial velocity 
curve is shown in Fig.\,\ref{fig:rvcurve}.

\setcounter{table}{2}
\begin{table}
\caption{Orbital parameters of HD\,11753.}
\label{tab:orbit}
\centering
\begin{tabular}{cr@{ $\pm$ }lc}
\hline
\hline

$P$ [d]                 & 1126.11   &  0.16  \\
$T_{\rm o}$          & 2453766.2  &  2.2    \\
$V_{\rm o}$ [km/s]      &   10.44   &  0.04  \\
$K_{\rm A}$ [km/s]      &    9.21   &  0.09  \\
$e$                     &    0.589  &  0.004 \\
$\omega$ [rad]		&    3.52   &  0.01  \\
$a\,\sin\,i$ [$R_\odot$]     &  165.4     &   2.2    \\
\hline
\end{tabular}
\end{table}

The orbital phase coverage of the new observations (FEROS, CORALIE, HARPS) 
has unfortunately been very poor. All of them were concentrated around the flat 
maximum, with the only exception one FEROS spectrum taken at $\phi$=0.03.
This fact hindered the identification of the period in recent works
(e.g., Paper\,1; Makaganiuk et al. \cite{maka12}) in spite of the large
number of available observations.

\begin{figure}
\centering
\includegraphics[angle=-90, width=0.4\textwidth]{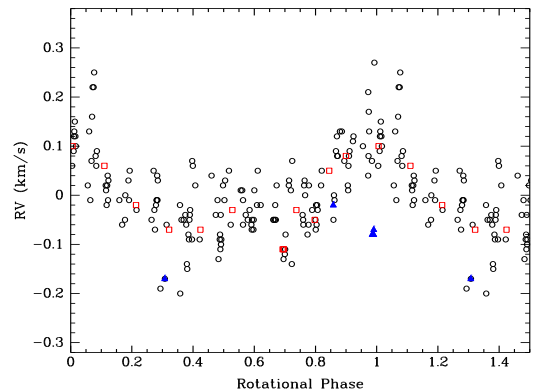}
\caption{Radial velocity residuals as a function of rotational phase. 
Symbols are as in Fig.\,\ref{fig:rvcurve}.}
\label{fig:residuals}
\end{figure}

All high-resolution observations fit the newly calculated orbit very well,
including the nine Palomar spectra taken by Dworetsky et al. (\cite{d82}) for 
which the root mean square (RMS) of residuals is 0.22 km/s. Residuals of 
CORALIE, FEROS, and HARPS measurements have an RMS of 84 m/s. Part of this 
dispersion is due to the line profile variability of HD\,11753. In fact, the 
residuals show a clear one-wave variation as a function of the rotational phase 
(see Fig.\,\ref{fig:residuals}), even though in cross-correlations we used a 
spectrum corresponding to solar chemical abundance as a template so that the 
main peculiar lines (e.g., those of \ion{Y}{ii}) have not contributed to our 
radial velocity measurements. After subtracting a soft curve to the residuals, 
we obtained an RMS of about 60 m/s, which can be considered as an estimate of 
the error of our CORALIE+FEROS+HARPS measurements. On the other hand, the 
Dworetsky et al. SAAO observations have RMS=1.4 km/s and Leone \& Catanzaro 
measurements RMS=0.5 km/s.
 
The calculated orbit is quite wide and eccentric. The orbital semiaxis is
probably around 700--750 R$\odot$, and the secondary star mass would be in the 
range 0.9--1.7 M$\odot$. The minimum values in these ranges were calculated 
from the radial velocity amplitude assuming that the primary star has a mass of
about 2.8~M$\odot$ (estimated from T$_{\rm eff}$=10\,600\,K, $\log g$=3.8). The 
minimum values were derived from the spectral lines of the companion star not 
being visible in the spectrum, and so the mass-ratio is probably below 0.6. 
Since the separation at periastron is of the order of 300~R$\odot$, no 
interaction is expected to take place between the companions, and the observed 
chemical asymmetries in the primary star surface are not related to binarity.

\section{Doppler imaging}

\subsection{Period determination}
\label{sec_period}

\begin{figure}
\centering
\includegraphics[angle=-90, width=0.45\textwidth]{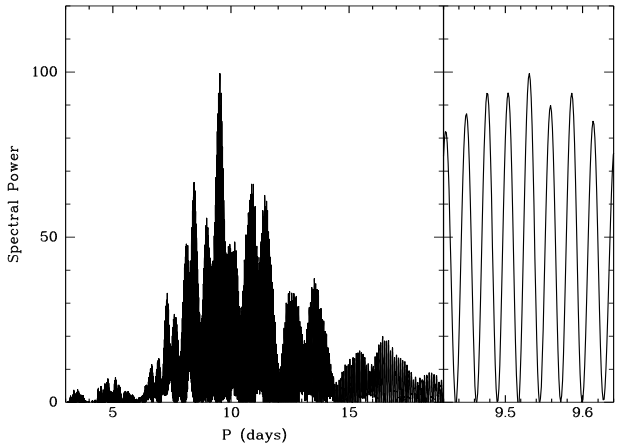}
\caption{Periodogram from the sum of the equivalent widths of seven \ion{Y}{ii}
lines (left panel) and zoom around the adopted period (right panel). }
\label{fig:pdg}
\end{figure}

All available datasets were used to determine the rotation period of HD~11753 
using equivalent width (EW) measurements of seven \ion{Y}{ii} lines: 
$\lambda\lambda$4309.62, 4374.94, 4398.02, 4682.32, 4900.13, 5084.73, and 
5662.95~{\AA}. Figure~\ref{fig:pdg} shows the periodogram calculated using the 
Lomb-Scargle method (Scargle \cite{scargle}). The long time span of the 
observations allows an accurate determination of the period; a least-square fit
of the sum of the EW of the seven  \ion{Y}{ii} lines with a cosine function 
yields $P=9.53077\pm0.00011$ days. However, owing to the long observational gap 
between the years 2000 and 2009, with only a few FEROS spectra in between, 
there are also other periods, separated by steps of 0.0273 days, that are 
compatible with the observations, as can be seen in the right-hand panel of 
Fig.\,\ref{fig:pdg}.

\begin{figure}
\centering
\includegraphics[width=0.4\textwidth]{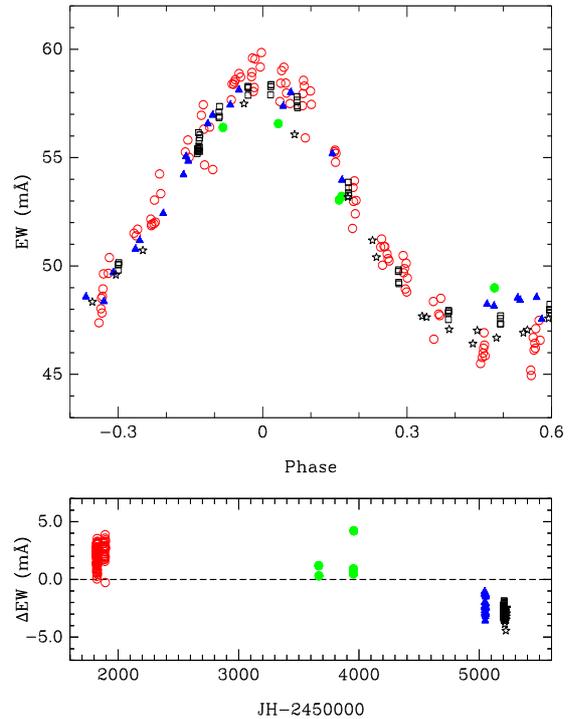}
\caption{Equivalent width of \ion{Y}{ii} lines. The average of seven 
\ion{Y}{ii} lines is plotted for different epochs: CORALIE 2000 (open 
circles), FEROS 2005-2006 (filled circles), CORALIE 2009 (triangles), HARPSpol 
2010 (squares), and CORALIE 2010 (stars). Lower panel: Secular variation of the
mean equivalent width after removing the rotational modulation. 
 }
\label{Ycurve}
\end{figure}

In addition to the rotational modulation of the line profiles, secular spectral
variability is detected. Specifically, the EW of all the \ion{Y}{ii} lines 
analysed here is approximately 10-20\% (4-6 m\AA) higher in the 2000 datasets 
than in the 2009-2010 datasets. The behaviour of the \ion{Y}{ii} line intensity
with phase is shown in Fig.\,\ref{Ycurve}, where a global correction of +0.48 
m{\AA} has been applied to 2009-2010 datasets in order to be able to plot the 
EW curves superimposed on each other. To show the secular variation more 
clearly, we removed the effect of rotational modulation using a least-square 
fit and plot the residual in the lower panel of Fig.\,\ref{Ycurve}. Parameters 
of the least-squares fit are listed in Table\,\ref{param_ycurves}. FEROS data 
are not included in this analysis because the number of datapoints is too small
for reliable determining the shape of the EW curve. Even though the behaviour 
of all the \ion{Y}{ii} lines is very similar, a small long term change in the 
amplitude of the variations can be seen. We note that this behaviour is not 
instrumental because it is not present, for example, in \ion{Ti}{ii} and 
\ion{Cr}{ii} lines (see Section~\ref{long_term}).

\begin{table*}
\caption{Secular variation of \ion{Y}{ii} curves. For the sum of the equivalent
widths of seven \ion{Y}{ii} lines, the mean value, amplitude, and time of 
maximum intensity is listed for five datasets.}
\label{param_ycurves}
\centering 
\begin{tabular}{l c c c c}
\hline\hline            
Dataset & mean JD & mean EW & EW semi-amplitude& T$_{\rm Ymax}$ \\ 
        &         &  m\AA   &  m\AA           &  \\
\hline    
Oct 2000  &2\,451\,823.6  &51.73$\pm$0.08   &6.49 $\pm$0.11   &2\,454\,009.497$\pm$0.026 \\
Dec 2000  &2\,451\,887.7  &52.18$\pm$0.16   &5.99 $\pm$0.22   &2\,454\,009.445$\pm$0.051 \\
Aug 2009  &2\,455\,050.0  &47.41$\pm$0.14   &5.29 $\pm$0.17   &2\,454\,009.552$\pm$0.053 \\
HARPS	  &2\,455\,207.3  &47.14$\pm$0.05   &5.59 $\pm$0.05   &2\,454\,009.491$\pm$0.014 \\
Jan 2010  &2\,455\,218.2  &46.46$\pm$0.08   &5.42 $\pm$0.13   &2\,454\,009.461$\pm$0.033 \\
\hline                       
\end{tabular}
\end{table*}

\begin{figure}
\centering
\includegraphics[width=0.4\textwidth]{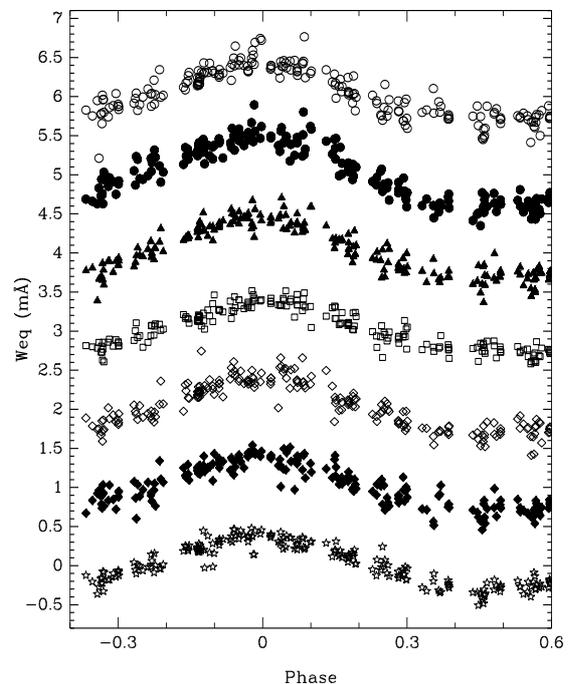}
\caption{Equivalent widths of seven \ion{Y}{ii} lines as a function of the 
rotational phase. From top to bottom: $\lambda\lambda$4309.62, 4374.94, 
4398.02, 4682.32, 4900.13, 5084.73, and 5662.95 {\AA.} }
\label{7curves}
\end{figure}

\begin{figure}
\centering
\includegraphics[width=0.2\textwidth]{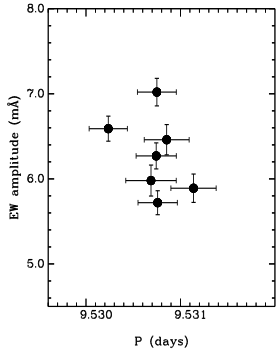}
\caption{Amplitude and period of the variations of seven \ion{Y}{ii} lines.
 }
\label{amp-P}
\end{figure}

To have an independent estimate of the period uncertainty we also calculated 
the period individually from the seven \ion{Y}{ii} lines used in the previous 
analysis by fitting cosine curves to the data. Curves for individual lines can 
be seen in Fig.\,\ref{7curves}, and the resulting parameters are 
plotted in Fig.\,\ref{amp-P}.

No significant evidence of any migration of surface spots is detected from the 
times of maximum \ion{Y}{ii} abundance. Naturally, since the rotational period 
has been determined from \ion{Y}{ii} abundance variations, global shifts in 
the phase cannot be detected (constant movement of the features would be 
interpreted as a different period), but abrupt movements of spots would still 
be measurable. 

In the following Doppler imaging analysis, the newly determined period of 
9.53077~days for the phase calculation is used, but we adopt the same T as was 
used in Paper\,1, namely 2\,451\,800.0 With these ephemeris the intensity of 
\ion{Y}{ii} lines is not maximum at $\phi=0.00$ as shown in 
Figs.\,\ref{Ycurve} and \ref{7curves}, but at the phase $\phi=0.8283$.

\subsection{Selection of spectral lines}

In Paper\,1 only one spectral line per element was used due to the possible 
radial stratification of the elements in the atmospheres of HgMn stars (e.g., 
Savanov \& Hubrig \cite{sav_hub}; Thiam et al. \cite{thiam}). Makaganiuk et al. 
(\cite{maka12}) have investigated possible stratification of \ion{Ti}{ii} and 
\ion{Y}{ii} in HD~11753 and find very marginal indication of stratification in 
\ion{Y}{ii}. For the current work, the line selection was based on the line
identifications in a slowly rotating HgMn star HD~175640 by Castelli \& Hubrig 
(\cite{castelli_hubrig}). Only non-blended lines that are detected in HD~11753
at a reasonable level are chosen for the analysis. For \ion{Sr}{ii} only one 
non-blended line is available, namely \ion{Sr}{ii} 4305.443~{\AA}. The 
$\log gf$ values are mainly obtained either from VALD (e.g., Piskunov et al. 
\cite{VALD1}; Kupka et al. \cite{VALD2}) or from Makaganiuk et al. 
(\cite{maka12}). In the case of \ion{Y}{ii} 5662.929~{\AA} a slight adjustment 
to the $\log gf$ value  was necessary to be able to fit all the \ion{Y}{ii} 
lines simultaneously (value used here is 0.450, instead of 0.160 given in VALD 
and 0.384 used by Makaganiuk et al. \cite{maka12}). The spectral lines used in 
this study and their parameters are given in Table~\ref{line_param}.

\begin{table}
\caption{Spectral line parameters used in this study. The ion, central 
wavelength, excitation energy, log gf value, and the source of log gf are given.
In the source M2012 means Makaganiuk et al. (\cite{maka12})}
\label{line_param}
\centering 
\begin{tabular}{c c c c c}
\hline\hline            
Ion & Wavelength [\AA] & Excit [eV] & log gf & source\\ 
\hline    
\ion{Y}{ii} & 4398.013 & 0.130 & -1.000 & VALD\\ 
\ion{Y}{ii} & 4900.120 & 1.033 & -0.090 & VALD\\
\ion{Y}{ii} & 5200.405 & 0.992 & -0.570 & VALD\\
\ion{Y}{ii} & 5662.929 & 1.944 &  0.450 & this work\\
\ion{Sr}{ii} & 4305.443 & 3.040 & -0.485 & M2012\\
\ion{Ti}{ii} & 4163.644 & 2.590 & -0.130 & VALD \\
\ion{Ti}{ii} & 4399.765 & 1.237 & -1.190 & VALD \\
\ion{Ti}{ii} & 4417.714 & 1.165 & -1.190 & VALD \\
\ion{Ti}{ii} & 4563.757 & 1.221 & -0.690 & VALD \\
\ion{Cr}{ii} & 4145.781 & 5.319 & -1.214 & M2012 \\
\ion{Cr}{ii} & 4275.567 & 3.858 & -1.730 & VALD \\
\ion{Cr}{ii} & 4554.988 & 4.071 & -1.438 & M2012 \\
\ion{Cr}{ii} & 4565.739 & 4.071 & -1.980 & VALD \\ 
\hline                       
\end{tabular}
\end{table}

\subsection{Surface distribution of chemical elements in 2000-2010}

We have obtained Doppler images of HD~11753 from CORALIE spectra for four 
different epochs. In Doppler imaging (see e.g., Vogt et al. \cite{vogt}; 
Piskunov et al. \cite{pisk90}), spectroscopic observations at different 
rotational phases are used to measure the rotationally modulated distortions in
the line profiles. In HD~11753 these distortions are produced by the 
inhomogeneous distribution of element abundance. Surface maps are constructed 
by combining all the observations from different phases and comparing them with 
synthetic model line profiles. For the inversion we used the INVERS7PD code 
originally written by Piskunov (see, e.g., Piskunov \cite{pisk91}) and modified
by Hackman et al. (\cite{hack}). This code is based on Tikhonov regularisation. 
The local line profiles were calculated using the same methods and codes as in 
Paper\,1, except for the line selection and line parameters, which were both 
discussed in detail in the previous section. The CORALIE data obtained in 2000
have already been used for Doppler imaging of \ion{Y}{ii}, \ion{Sr}{ii}, and 
\ion{Ti}{ii} lines in Paper\,1, but then using the period determination based 
on 2000 data alone, and using one spectral line per element. Here, the new 
period determination is adopted and, in most cases, four spectral lines have 
been used simultaneously in the inversion to obtain the map of that element. 
In addition, the regularisation parameter used in the inversion was changed to 
account for the weaker variability of \ion{Ti}{ii} and \ion{Cr}{ii} lines and 
to make the maps fully compatible with the maps of Makaganiuk et al. 
(\cite{maka12}). The August 2009 and January 2010 maps are previously 
unpublished, but some preliminary results have been presented in Korhonen et 
al. (\cite{kor_IAUS273}).

As in Paper\,1 we use v$\sin i$=13.5 km/s for the inversions, except in the 
case of \ion{Sr}{ii} where 12.5 km/s provides a much better fit. We run 
tests using the high-resolution HARPSpol observations, and the results clearly 
show that the best fit to the \ion{Y}{ii}, \ion{Ti}{ii}, and \ion{Cr}{ii} in the
inversion process is obtained using vsini 13.5 km/s, whereas \ion{Sr}{ii} 
requires smaller vsini. Similar difference in determined rotational velocity 
has been seen before in HgMn stars, at least for \ion{Hg}{ii} in HD~158704 
(Hubrig et al. \cite{hubrig99}. Makaganiuk et al. (\cite{maka12}) determine an
inclination $i=65.7^{\circ}\pm 7.1^{\circ}$ for HD~11753. We adopt here the 
inclination of $53^{\circ}$ that was used in Paper\,1. This lower inclination 
gives us a better fit than the higher value used by Makaganiuk et al. 
(\cite{maka12}).

\begin{figure}
\centering
\includegraphics[width=0.49\textwidth]{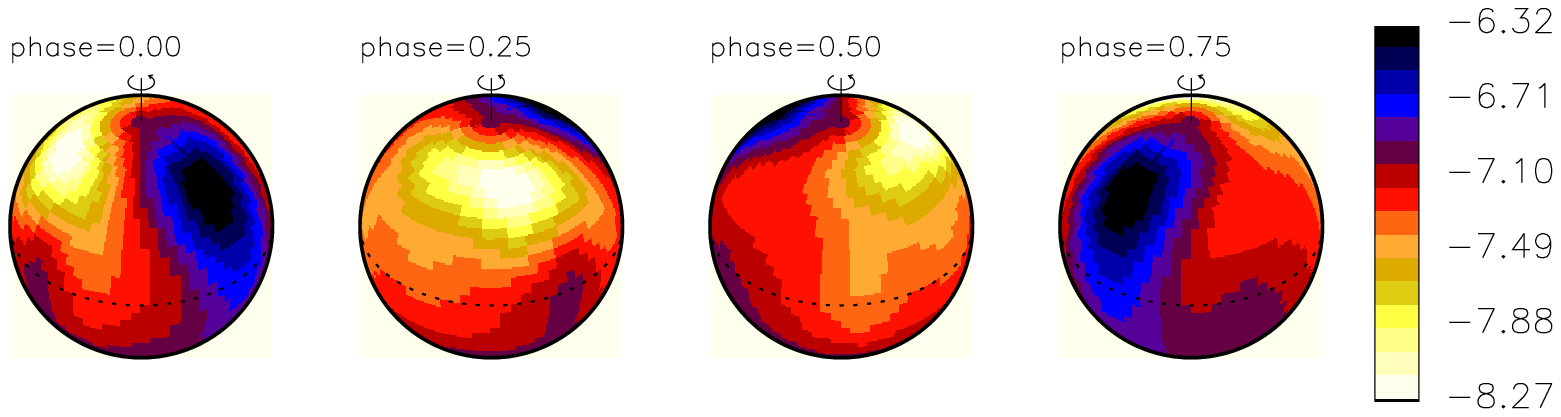}
\includegraphics[width=0.49\textwidth]{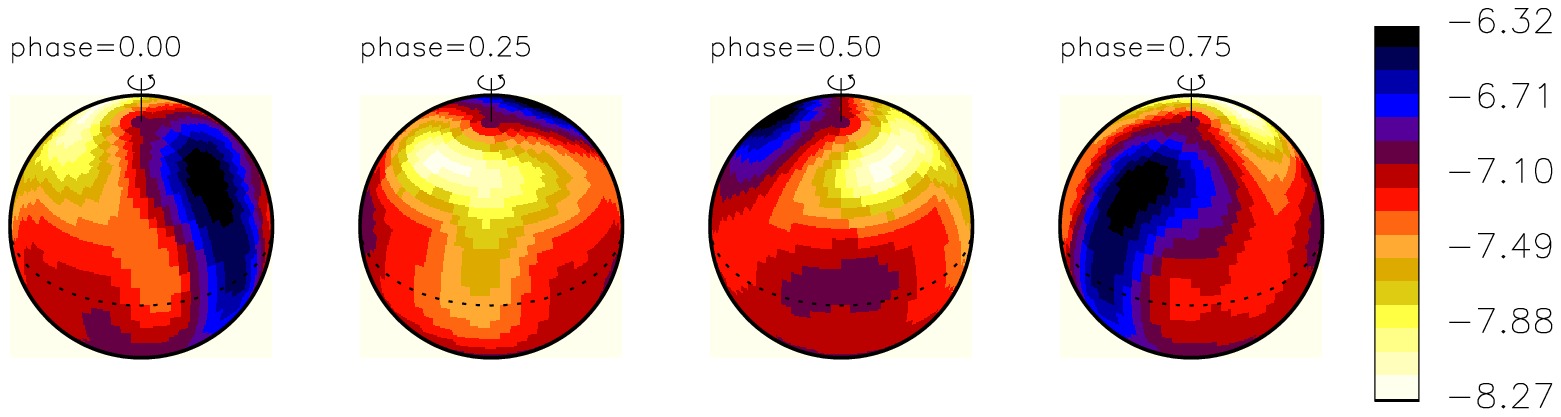}
\includegraphics[width=0.49\textwidth]{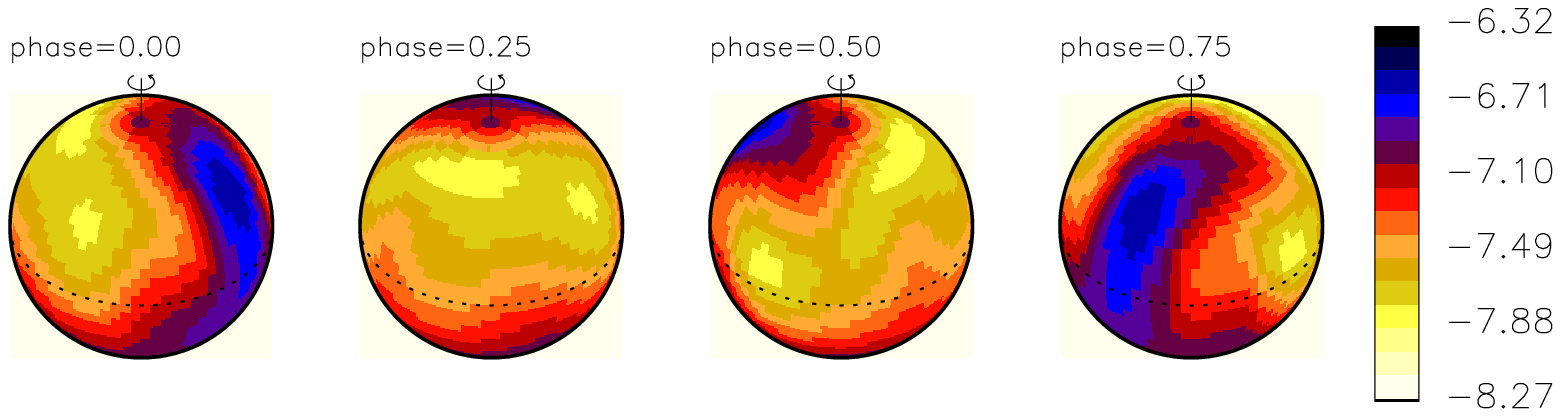}
\includegraphics[width=0.49\textwidth]{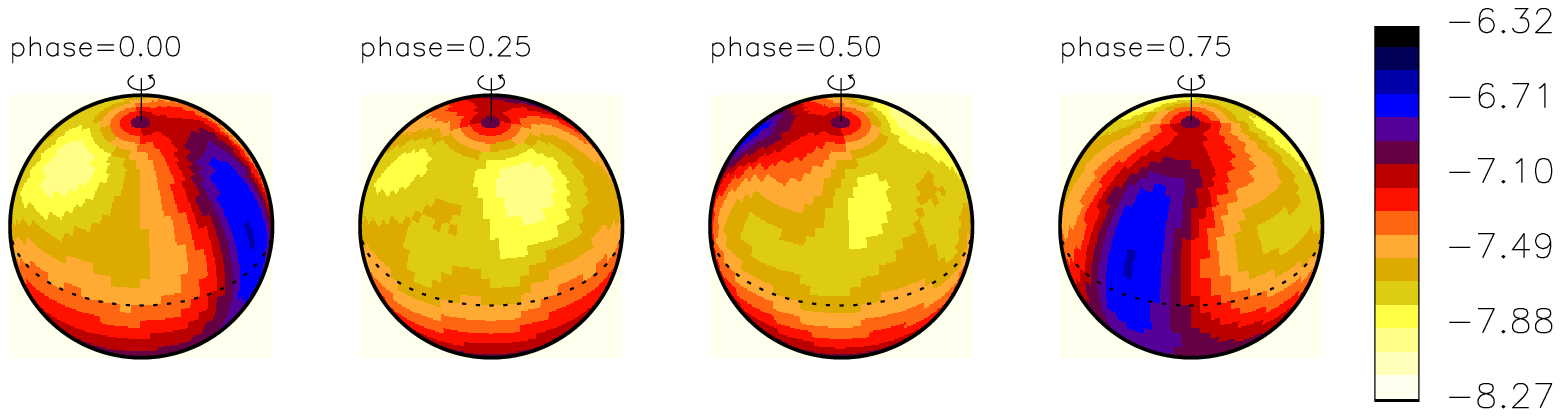}
\caption{Doppler maps of HD~11753 obtained from CORALIE data simultaneously 
  from \ion{Y}{ii} 4398.013~{\AA}, 4900.120~{\AA}, 5200.405~{\AA}, and 
  5662.929~{\AA} lines. Epochs of the maps are from top to bottom: October 
  2000, December 2000, August 2009, and January 2010. The surface distribution 
  is shown for four different phases 0.25 apart. The maps from the year 2000 
  have already been published in Paper\,1, but here they are shown using four 
  \ion{Y}{ii} lines simultaneously, using a refined regularisation parameter, 
  and the newly determined period. The abundance scale for different 
  epochs is the same.}
\label{Ymaps}
\end{figure}

\begin{figure}
\centering
\includegraphics[width=0.49\textwidth]{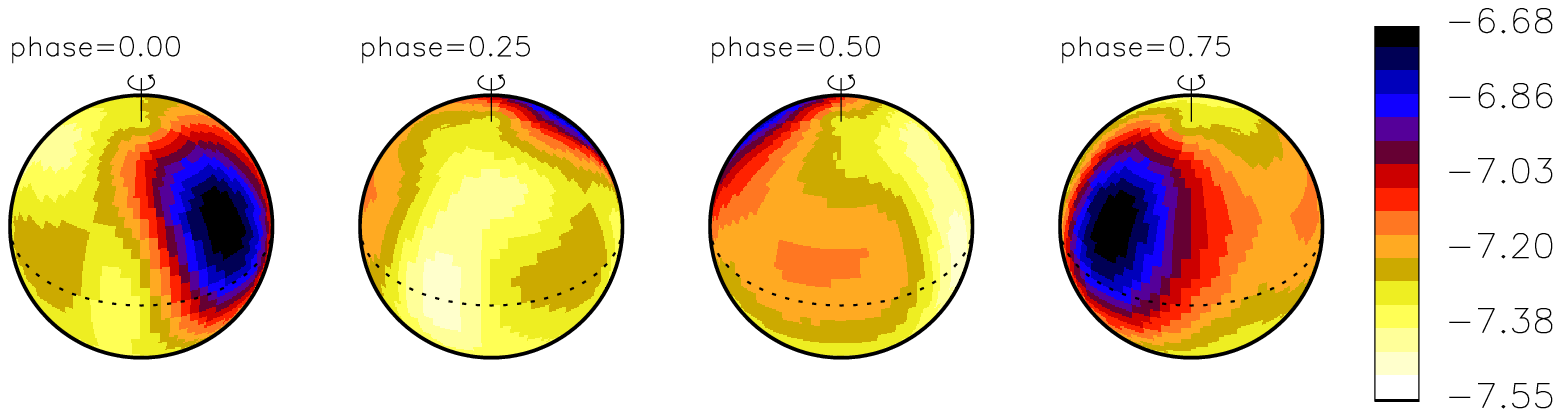}
\includegraphics[width=0.49\textwidth]{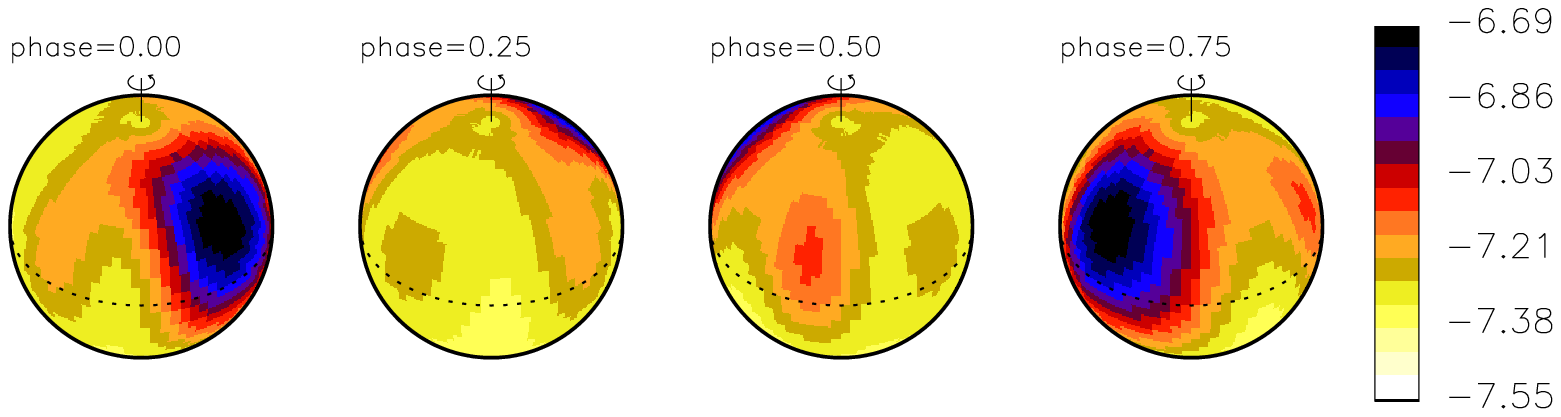}
\includegraphics[width=0.49\textwidth]{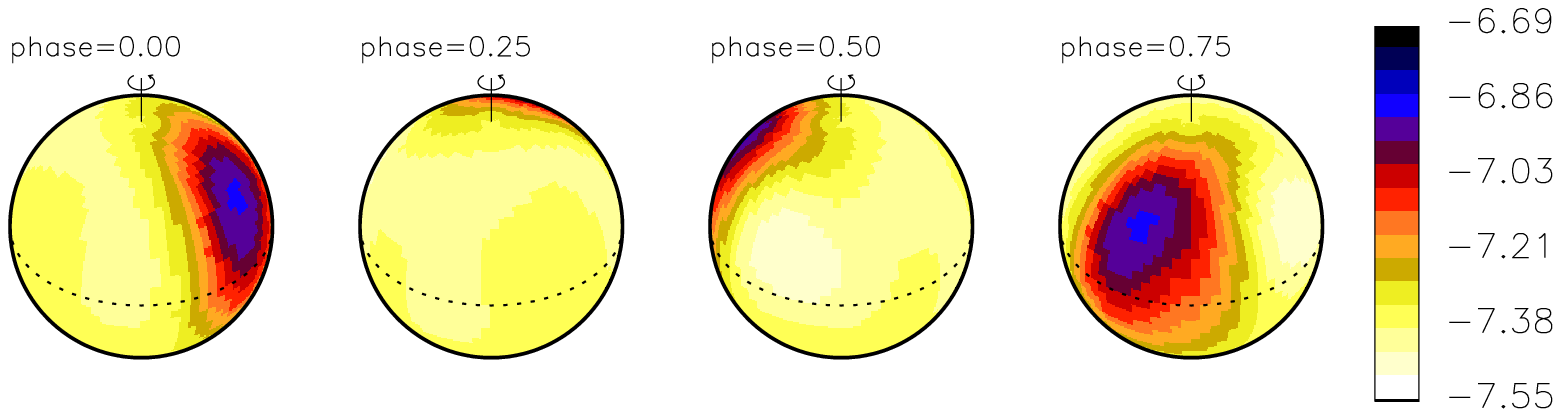}
\includegraphics[width=0.49\textwidth]{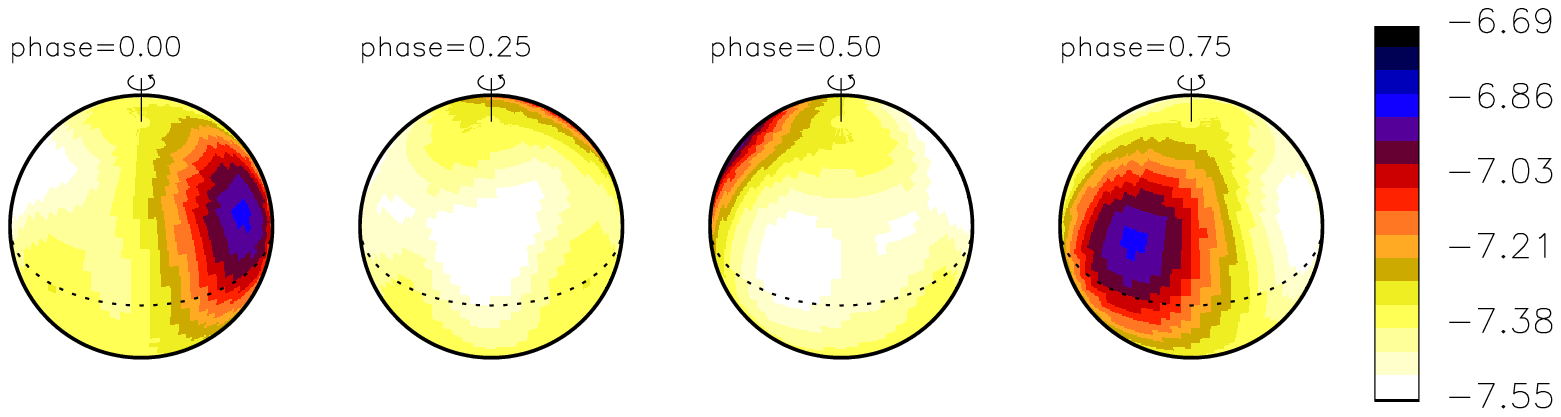}
\caption{Same as in Fig.~\ref{Ymaps}, but now for \ion{Sr}{ii} line 
4305.443~{\AA}.}
\label{Srmaps}
\end{figure}

\begin{figure}
\centering
\includegraphics[width=0.49\textwidth]{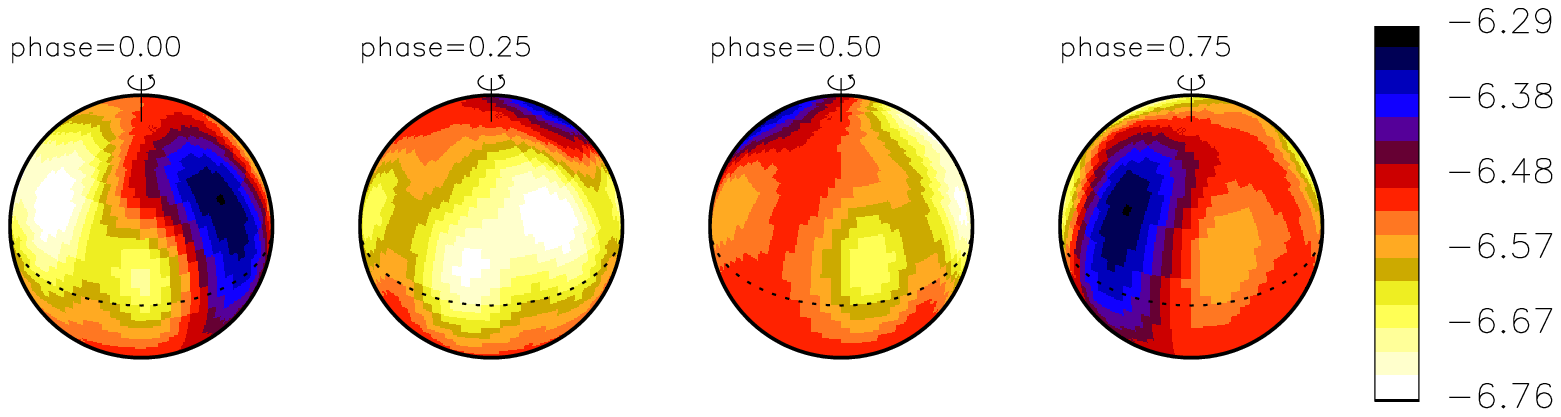}
\includegraphics[width=0.49\textwidth]{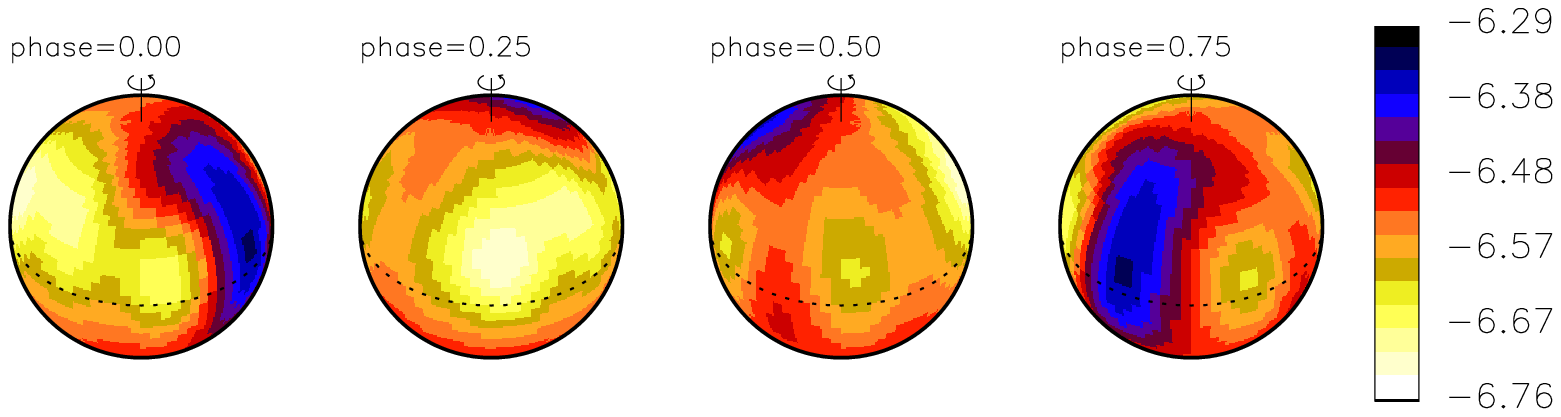}
\includegraphics[width=0.49\textwidth]{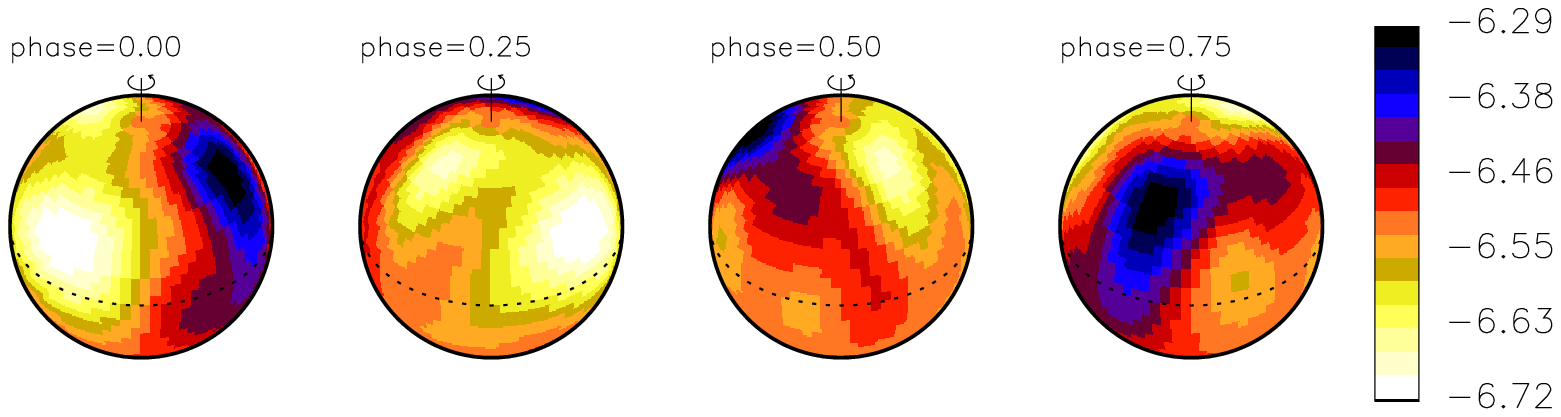}
\includegraphics[width=0.49\textwidth]{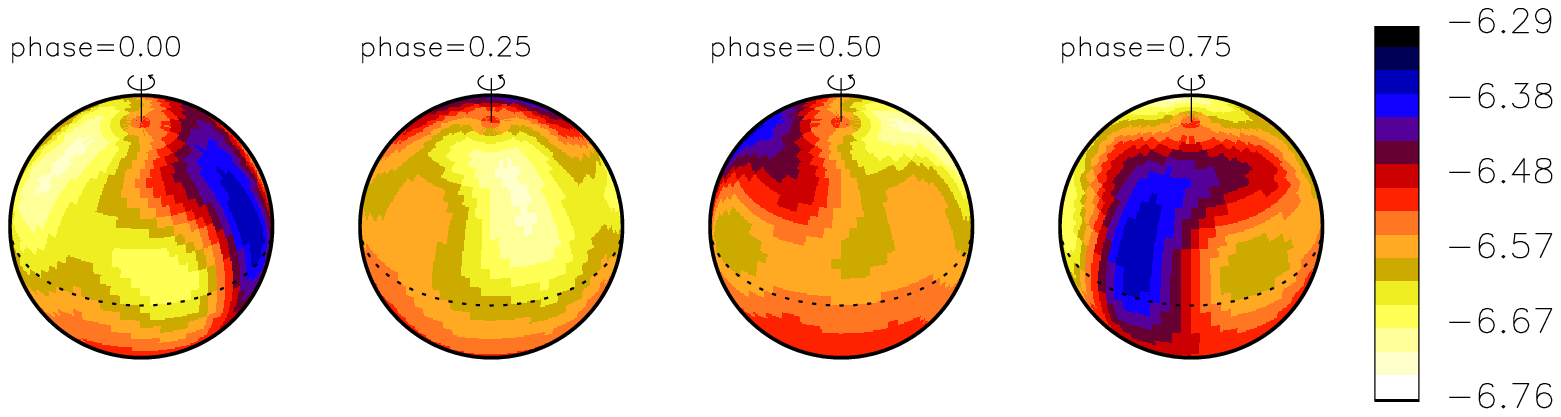}
\caption{Same as in Fig.~\ref{Ymaps}, but now for \ion{Ti}{ii} lines 
  4163.644~{\AA}, 4399.765~{\AA}, 4417.714~{\AA}, and 4563.757~{\AA} used 
  simultaneously.}
\label{Timaps}
\end{figure}

\begin{figure}
\centering
\includegraphics[width=0.49\textwidth]{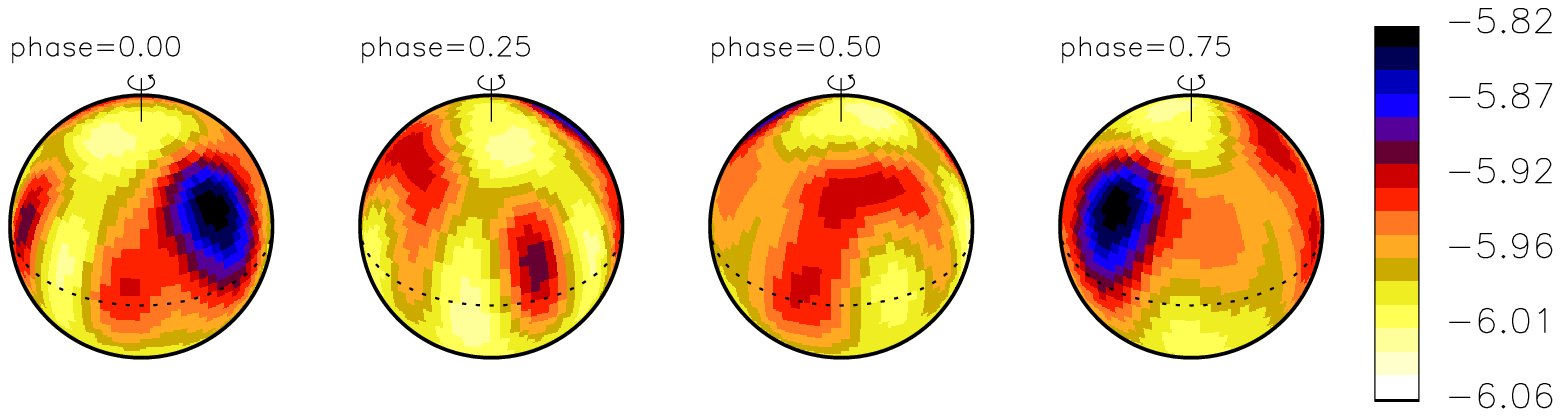}
\includegraphics[width=0.49\textwidth]{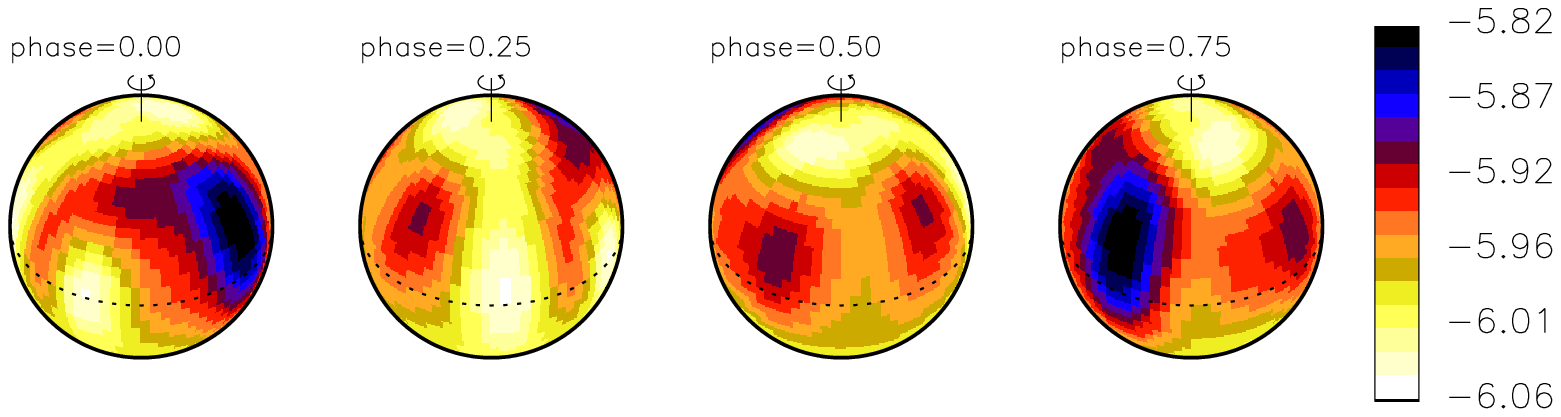}
\includegraphics[width=0.49\textwidth]{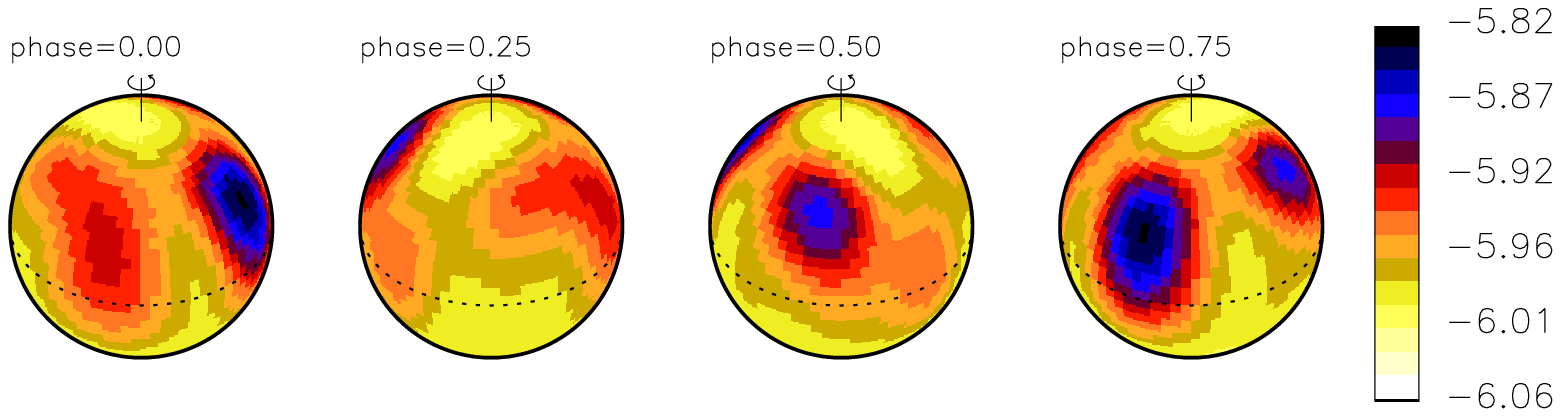}
\includegraphics[width=0.49\textwidth]{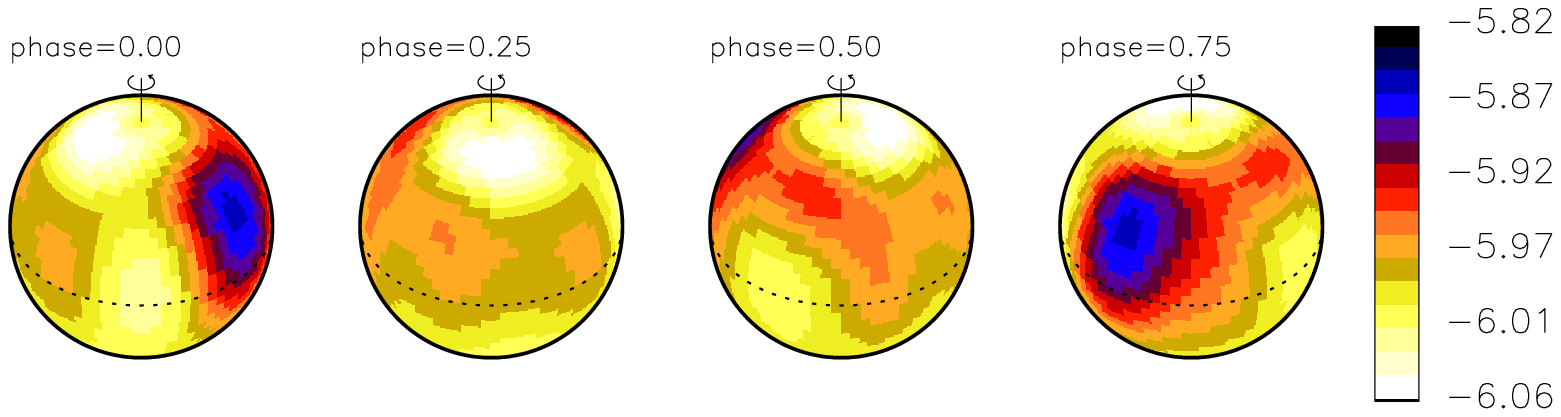}
\caption{Same as in Fig.~\ref{Ymaps}, but now for \ion{Cr}{ii} lines 
  4145.781~{\AA}, 4275.567~{\AA}, 4554.988~{\AA}, and 4565.739~{\AA} used 
  simultaneously.}
\label{Crmaps}
\end{figure}

\onlfig{12}{
\begin{figure}
\centering
\includegraphics[width=0.42\textwidth]{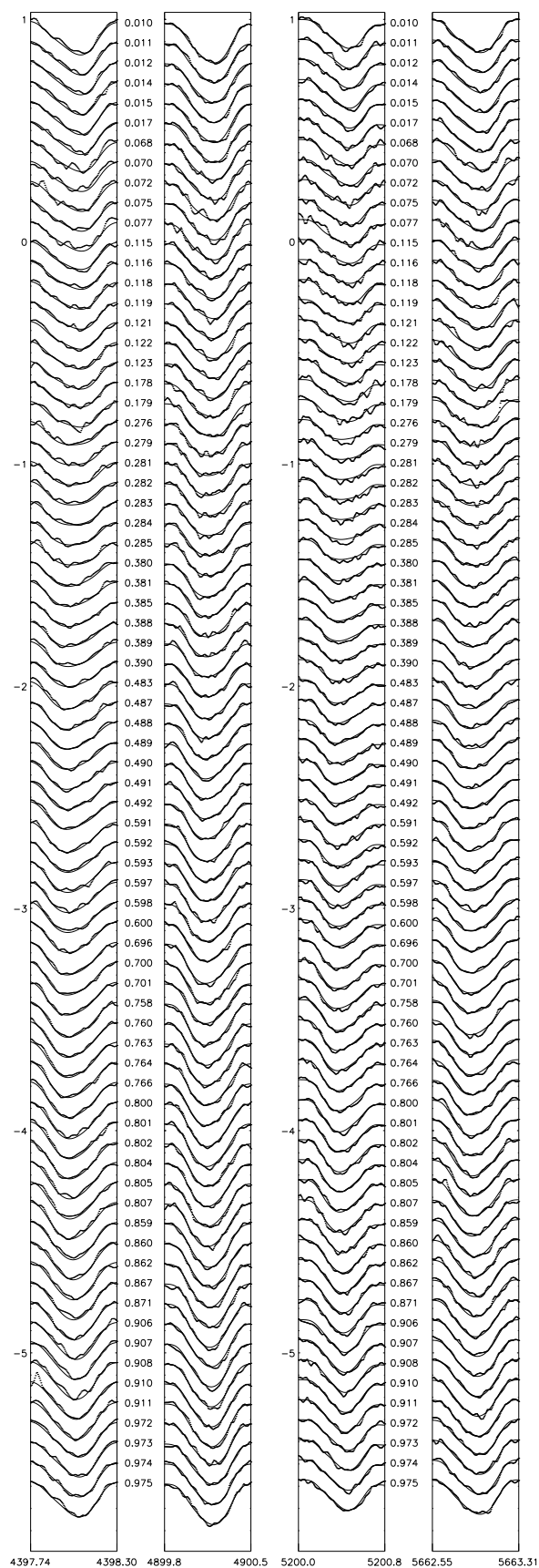}
\caption{October 2000 CORALIE observations (thin line) together with the model 
fit (thick line) for \ion{Y}{ii} lines used in Doppler imaging (map in 
Fig.~\ref{Ymaps}).}
\label{set1Yspectra}
\end{figure}}

\onlfig{13}{
\begin{figure}
\centering
\includegraphics[width=0.49\textwidth]{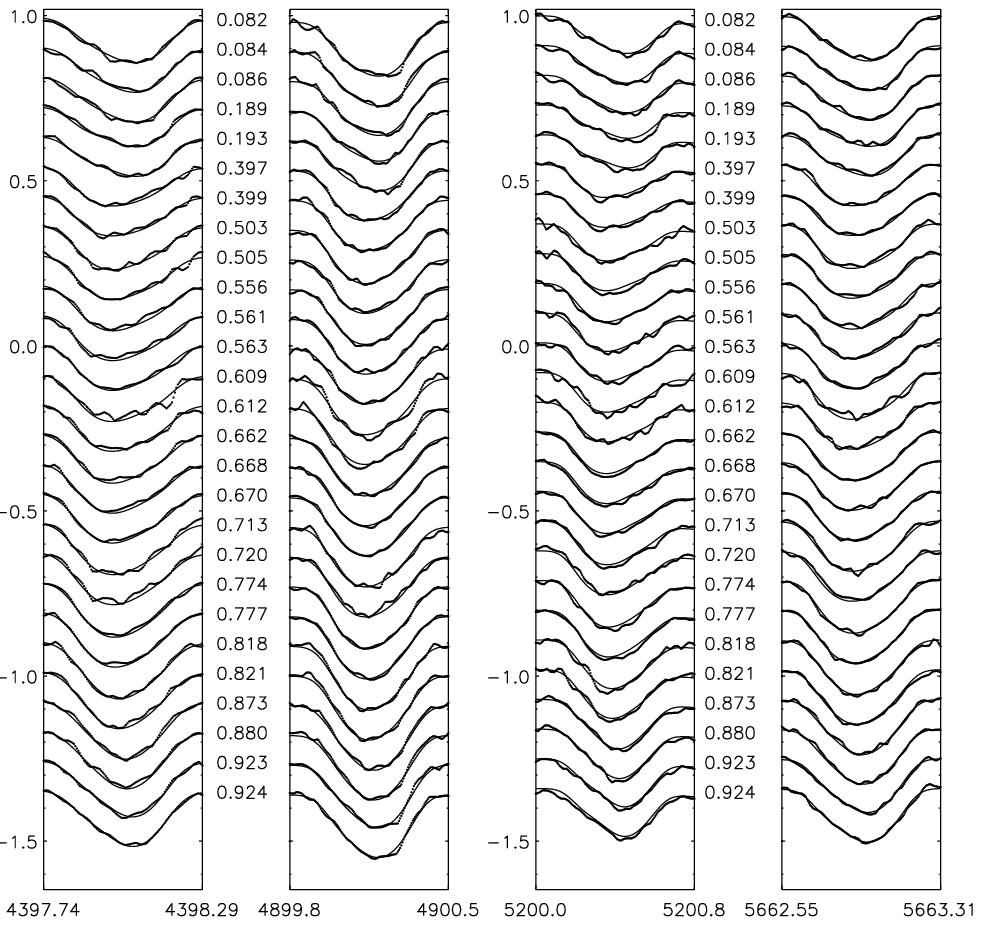}
\caption{Same as on-line Fig. \ref{set1Yspectra} except now for December 2000 CORALIE \ion{Y}{ii} observations.}
\label{set2Yspectra}
\end{figure}}

\onlfig{14}{
\begin{figure}
\centering
\includegraphics[width=0.49\textwidth]{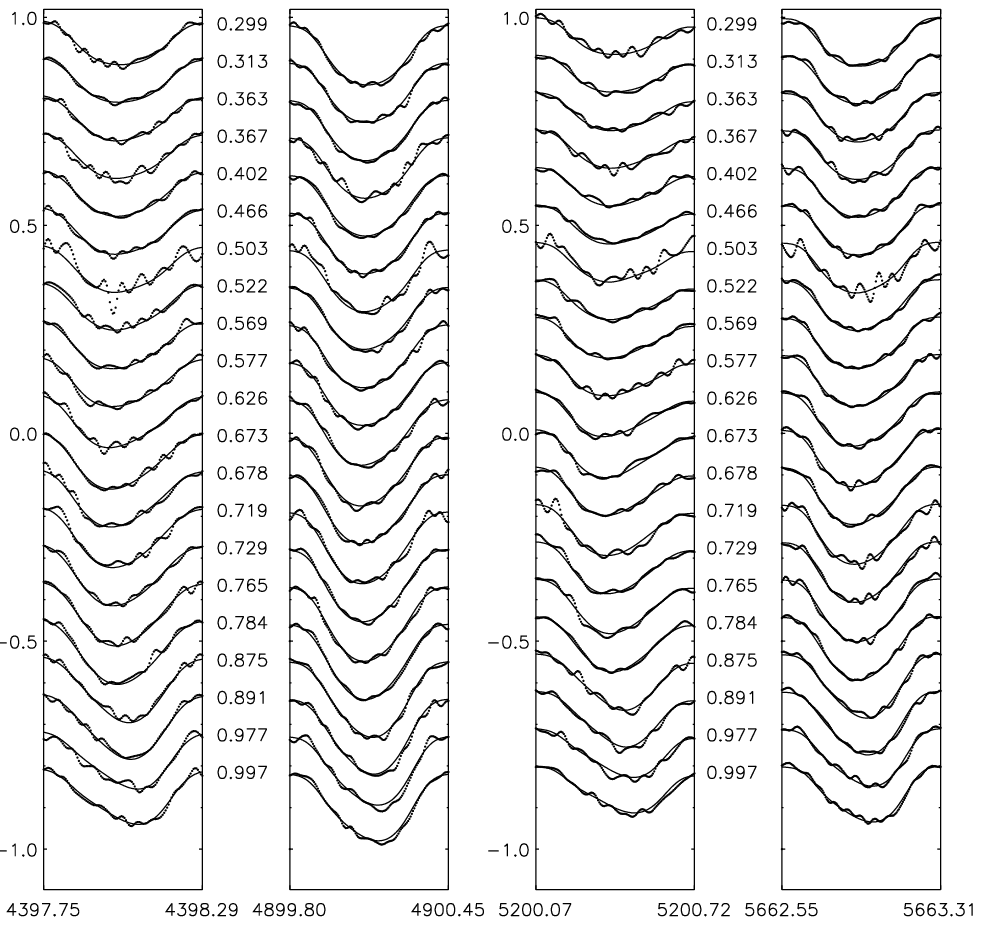}
\caption{Same as on-line Fig. \ref{set1Yspectra} except now for August 2009 CORALIE \ion{Y}{ii} observations.}
\label{2009Yspectra}
\end{figure}}

\onlfig{15}{
\begin{figure}
\centering
\includegraphics[width=0.49\textwidth]{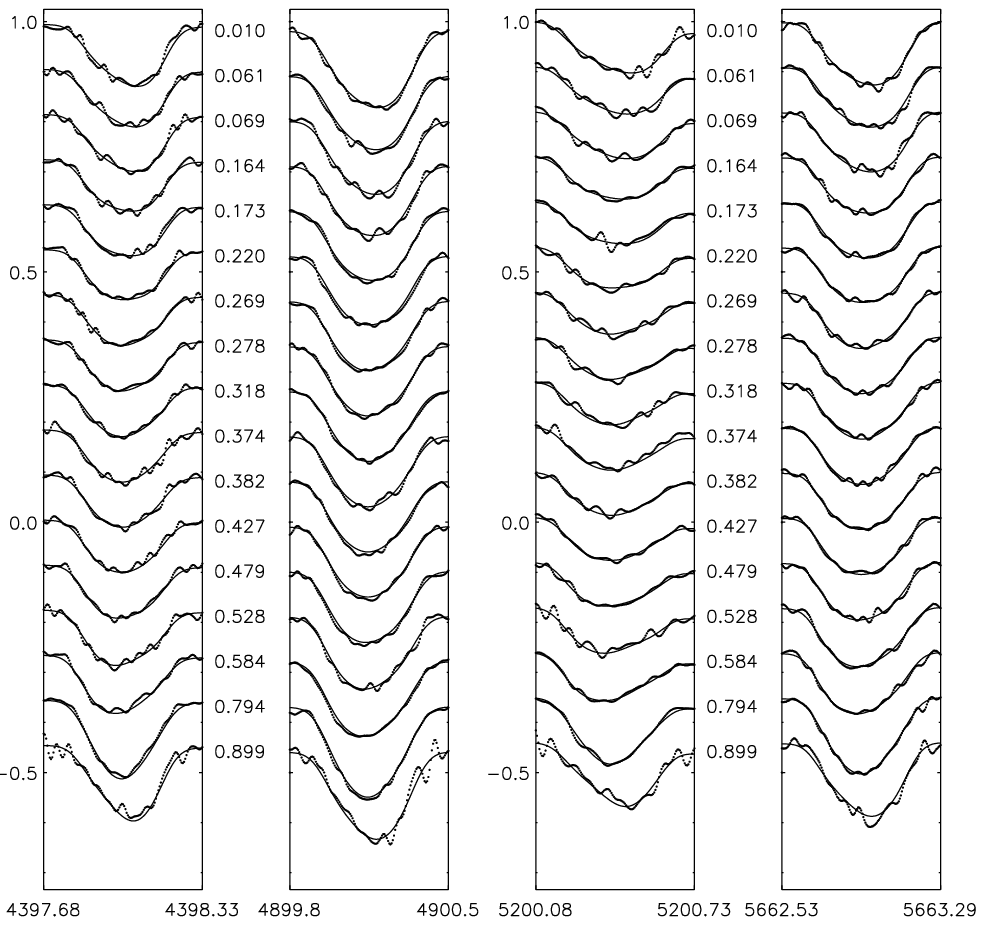}
\caption{Same as on-line Fig. \ref{set1Yspectra} except now for January 2010 CORALIE \ion{Y}{ii} observations.}
\label{2010Yspectra}
\end{figure}}

\onlfig{16}{
\begin{figure}
\centering
\includegraphics[width=0.42\textwidth]{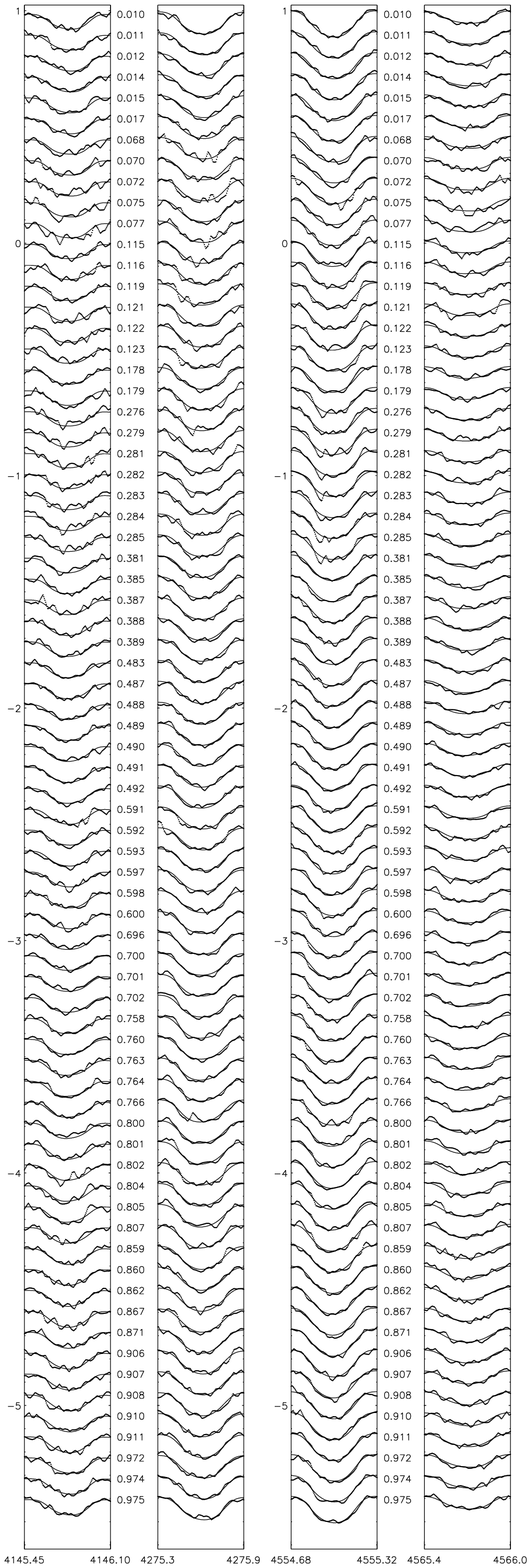}
\caption{Same as on-line Fig. \ref{set1Yspectra} except now for October 2000 CORALIE \ion{Cr}{ii} observations.}
\label{set1Crspectra}
\end{figure}}

\onlfig{17}{
\begin{figure}
\centering
\includegraphics[width=0.49\textwidth]{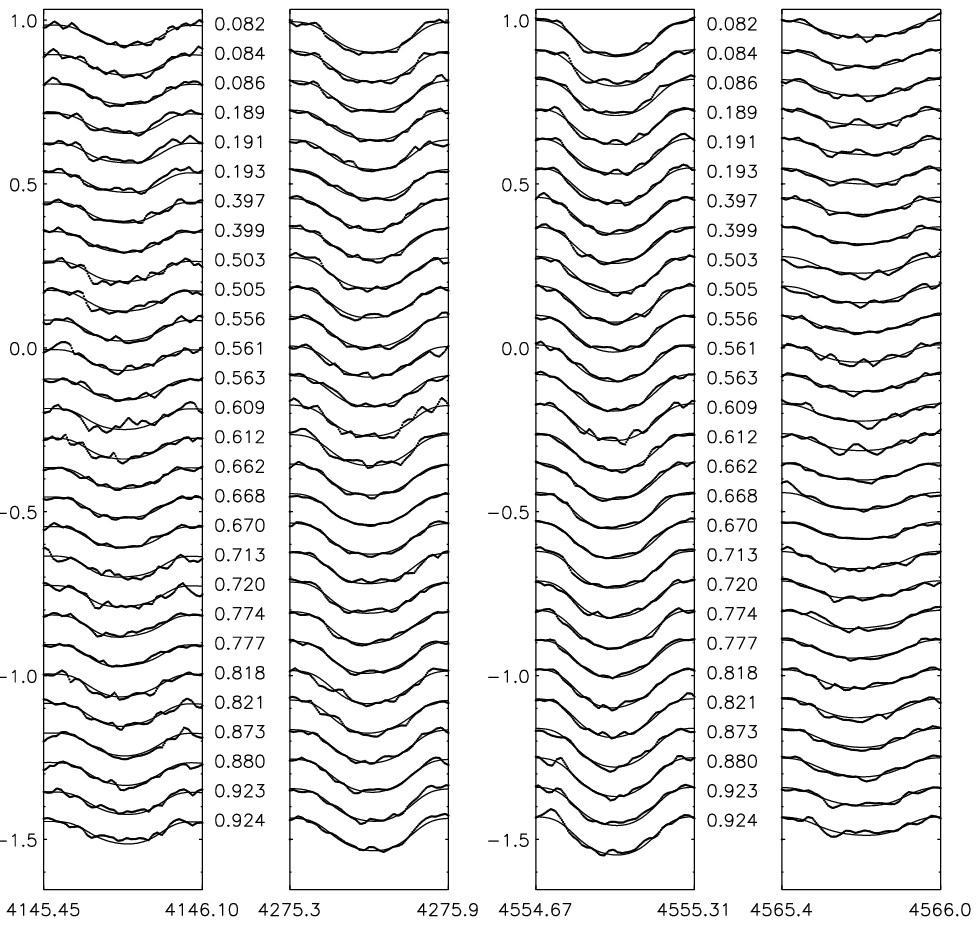}
\caption{Same as on-line Fig. \ref{set1Yspectra} except now for December 2000 CORALIE \ion{Cr}{ii} observations.}
\label{set2Crspectra}
\end{figure}}

\onlfig{18}{
\begin{figure}
\centering
\includegraphics[width=0.49\textwidth]{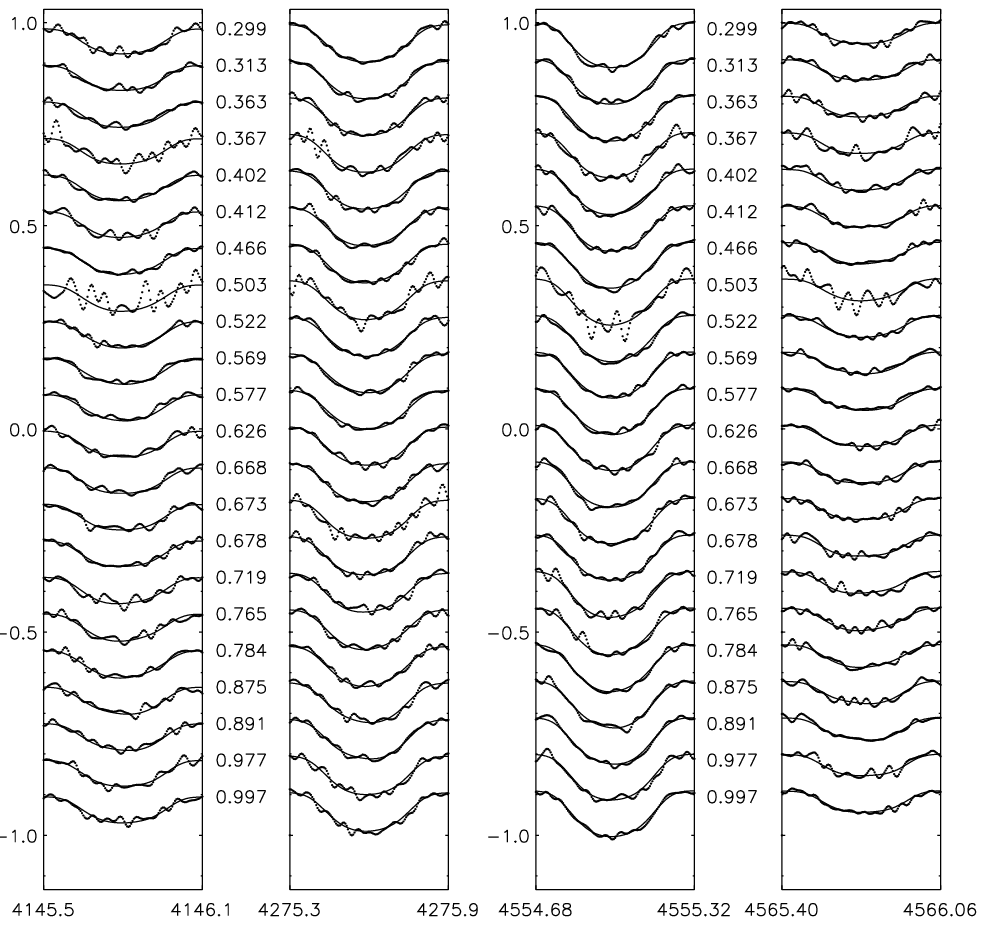}
\caption{Same as on-line Fig. \ref{set1Yspectra} except now for August 2009 CORALIE \ion{Cr}{ii} observations.}
\label{2009Crspectra}
\end{figure}}

\onlfig{19}{
\begin{figure}
\centering
\includegraphics[width=0.49\textwidth]{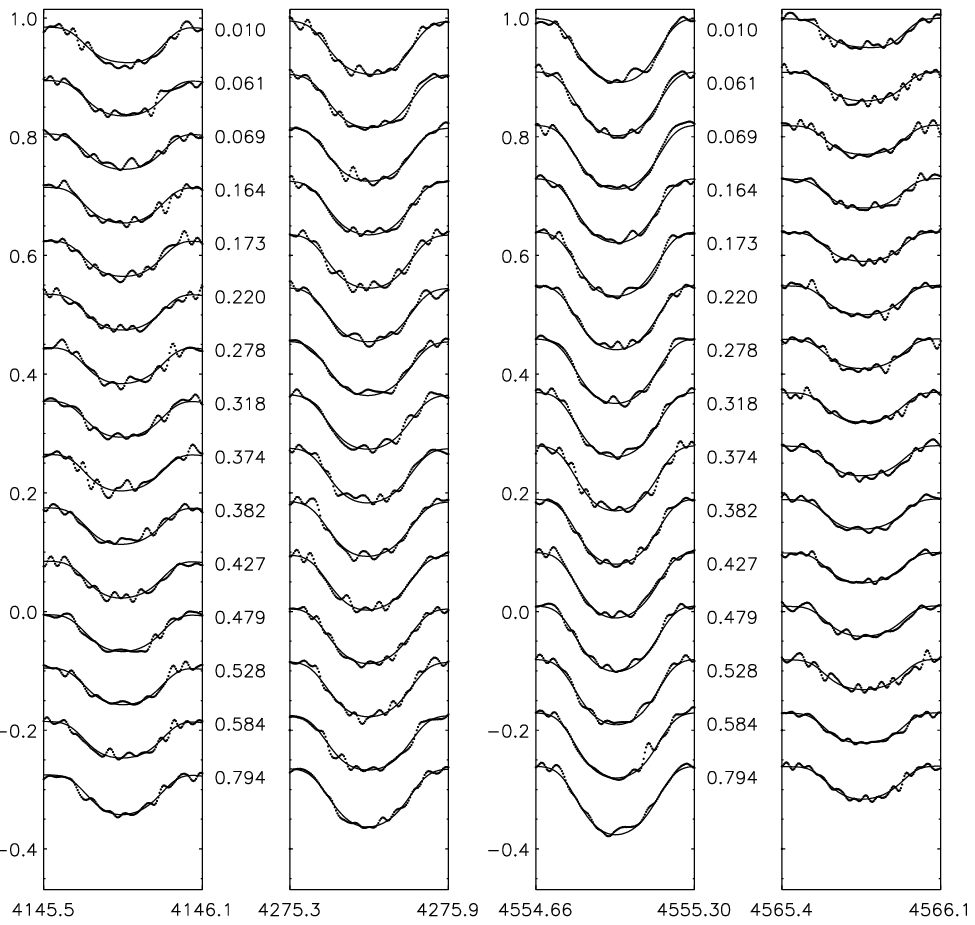}
\caption{Same as on-line Fig. \ref{set1Yspectra} except now for January 2010 CORALIE \ion{Cr}{ii} observations.}
\label{2010Crspectra}
\end{figure}}

The \ion{Y}{ii} distributions for the four datasets are shown in 
Fig.~\ref{Ymaps}. The maps for all the four epochs (October 2000, December 
2000, August 2009, and January 2010) show a high-abundance spot at the phases 
0.75--1.0 extending from the equatorial region all the way to the visible 
pole. Similarly in all the maps there is a lower abundance region around phases 
0.2--0.4. Since the rotation period was determined using EW of \ion{Y}{ii} 
lines its abundance pattern is not expected to show any global longitude drift.

The \ion{Sr}{ii} distribution is recovered using only the line at 
4305.443~{\AA}, because no other unblended \ion{Sr}{ii} lines were available 
for the analysis. The results for the four epochs are shown in 
Fig.~\ref{Srmaps}. All the maps show a very similar spot configuration with a 
strong high-abundance spot around the phases 0.75 to 1.0, and the rest of the 
stellar surface having a relatively low \ion{Sr}{ii} abundance. The exact 
extent and configuration of the high-abundance regions change from epoch to 
epoch.

Figure \ref{Timaps} shows the \ion{Ti}{ii} distributions for the four epochs. 
The spot distribution shows a main high-abundance spot around the phases 
0.75--1.00, and patches of high and lower abundance elsewhere on the stellar 
surface. The lowest abundance is concentrated around the phases 0.00 to 0.30. 

In Fig. \ref{Crmaps} the \ion{Cr}{ii} distribution is shown for the four 
epochs. Again the highest abundances occur around the phases 0.75 to 1.00. The 
polar regions, on the other hand, always have low abundance, and the equatorial 
region is dominated by a high-abundance spot ring. 

The observed spectral line profiles and the model fits to them are shown in 
the on-line Figs.~\ref{set1Yspectra}--\ref{2010Yspectra} for \ion{Y}{ii} and 
Figs.~\ref{set1Crspectra}--\ref{2010Crspectra} for \ion{Cr}{ii}. These two 
elements were chosen because examples as they show the strongest and weakest 
variability.

On the whole, at all epochs the main elemental spots retain quite stably their 
positions on the stellar surface the almost ten-year period our observations 
cover. The exact shape of the spots changes, though. Also, in all the maps, for
different epochs and elements, all the abundances are higher than the solar 
abundance of that element. 

In Hubrig et al.~(\cite{Hubrig_mg}) detailed analysis of the magnetic field in 
HD~11753 is carried out using HARPSpol data of January 2010. Their results
show that there seems to be a correlation between the elemental spots and 
magnetic fields and their polarities. \ion{Y}{ii} and \ion{Ti}{ii} lines 
reveal a weak negative magnetic field at the rotational phase 0.2 with 
3$\sigma$ significance. On the other hand, \ion{Cr}{ii} and \ion{Fe}{ii} show 
a weak positive magnetic field at phase 0.78, again with 3$\sigma$ 
significance. These phases are exactly where our maps from January 2010 show 
the prominent elemental spots, with the main high abundance spot around phase 
0.75 and the main lower abundance spot at phases 0.2 to 0.4. Hubrig et al. 
(\cite{Hubrig_mg}) find similar correlation between magnetic field polarity and
low or high abundance spots also in other HgMn stars. Additional magnetic field 
measurements of these targets are needed to confirm these findings.

\subsection{Comparison to the published maps}
\label{comp}

\begin{figure}
\centering
\includegraphics[width=0.49\textwidth]{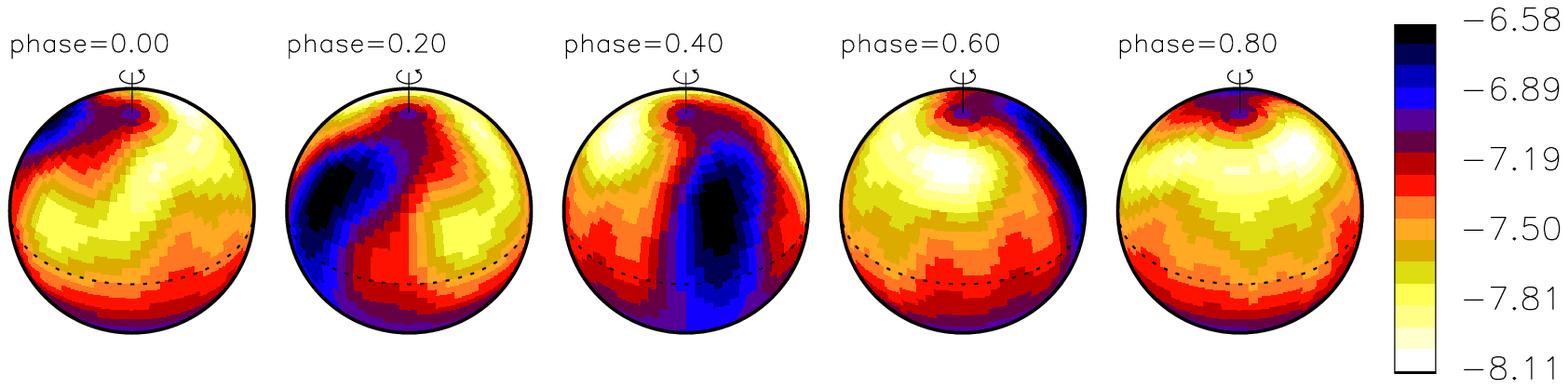}
\includegraphics[width=0.49\textwidth]{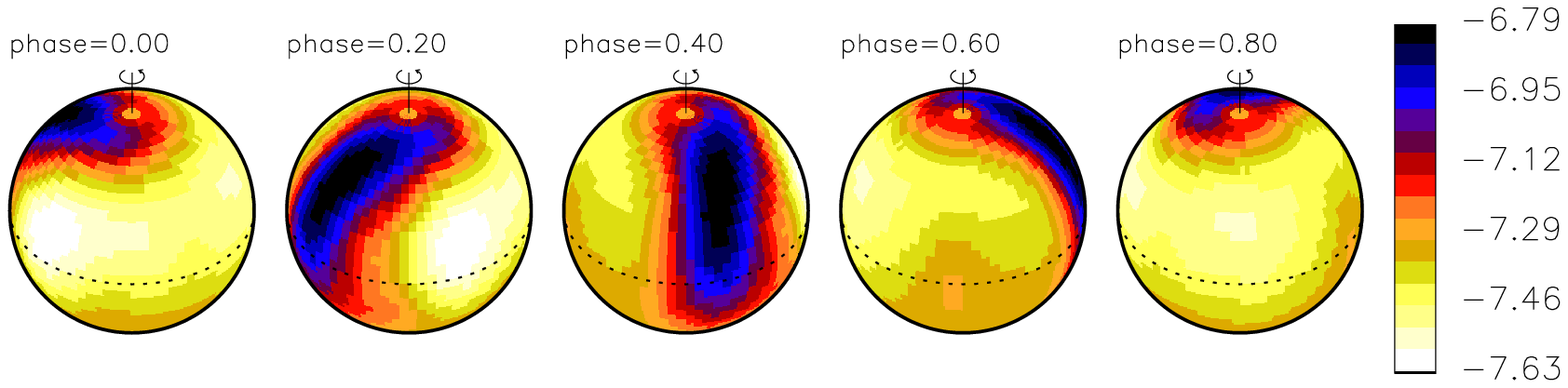}
\includegraphics[width=0.49\textwidth]{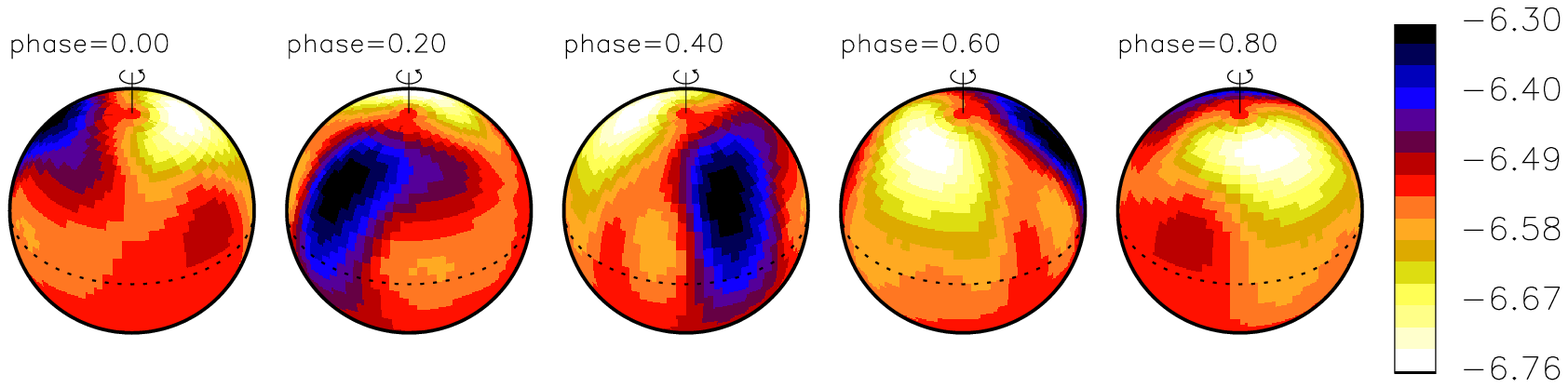}
\includegraphics[width=0.49\textwidth]{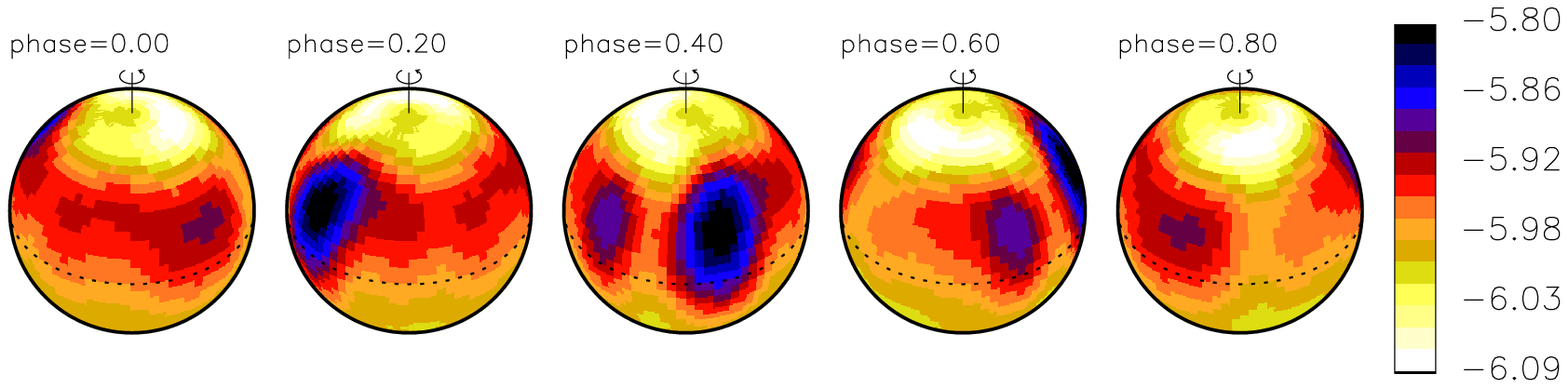}
\caption{Doppler maps of HD~11753 from January 2010 data obtained with 
  HARPSpol. The maps are from top to bottom: \ion{Y}{ii}, \ion{Sr}{ii}, 
  \ion{Ti}{ii} and \ion{Cr}{ii}. These same data have been used for Doppler 
  imaging by Makaganiuk et al. (\cite{maka12}). Here, same lines and 
  inclination angle as for the CORALIE datasets have been used, but now the 
  ephemeris is from Makaganiuk et al. (\cite{maka12}) to enable an easy 
  comparison. The results are presented from the same phases as used by
    Makaganiuk et al. (\cite{maka12}), and are virtually identical to theirs.}
\label{HARPS_maps}
\end{figure}

To test the reliability of our methods and codes, we also obtained elemental 
maps from the HARPSpol dataset, which was used by Makaganiuk et al. 
(\cite{maka12}) for obtaining elemental maps for \ion{Y}{ii}, \ion{Sr}{ii}, 
\ion{Ti}{ii}, and \ion{Cr}{ii}. We compared our results to their published 
maps. The core of the code used by Makaganiuk et al., INVERS10, is the same as 
in the code used here. Both codes have gone through several developments since 
the common version, INVERS7. The main differences are in the minimisation and 
coordinate system, where INVERS10 uses an equal surface area grid and INVERS7PD
uses a cartesian coordinate system. Also the local line profiles are calculated
differently in the two works. Here, the free available SPECTRUM code (Gray \& 
Corbally \cite{SPECTRUM}) has been used.

For this comparison we used the lines and line parameters given in Table 
\ref{line_param} and the inclination of 53$^{\circ}$, as is also used for our 
CORALIE data. On the other hand, the ephemeris used by Makaganiuk et al. 
(\cite{maka12}) is different from ours, so for an easier comparison we have 
used their ephemeris for the HARPSpol data presented in Fig.~\ref{HARPS_maps}. 
The results are virtually identical with the earlier published ones. The only 
slight difference is in the \ion{Cr}{ii} map, where the low abundance spot at 
the phase 0.5 is more pronounced in our map than in the one by Makaganiuk et 
al. (\cite{maka12}).

\subsection{Comparison of the maps obtained from 2010 January CORALIE and 
HARPSpol data}

A target with a low $v\sin i$, like HD~11753, would ideally require 
observations from an instrument with a very high resolving power, like 
HARPSpol. CORALIE's resolving power of $\sim$55,000 is relatively low, and 
therefore it is important to establish what effect that has on the resulting 
maps.

The 2010 January CORALIE dataset and 2010 January HARPSpol data were basically 
obtained immediately after each other with the HARPSpol dataset spanning the 
first half of January and the CORALIE dataset the second half. Here we compare 
maps obtained from these two datasets, which should not really show differences
in spot configurations, but are obtained with different instruments. HARPSpol 
S/N and resolution are superior to the CORALIE one, therefore providing a 
crucial test of the usability of CORALIE observations for the Doppler imaging 
of HD~11753.

\begin{figure}
\centering
\includegraphics[width=0.242\textwidth]{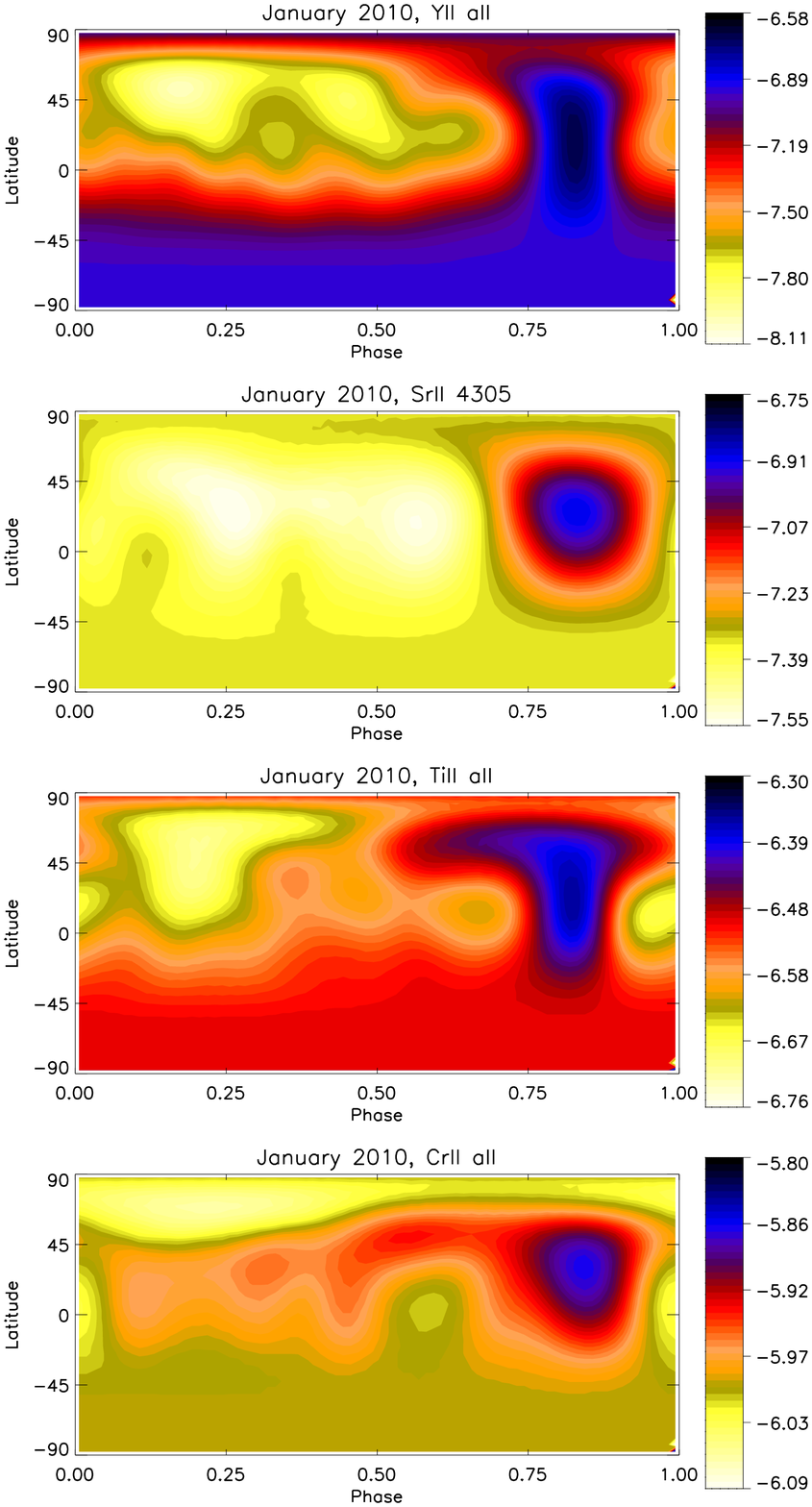}
\includegraphics[width=0.242\textwidth]{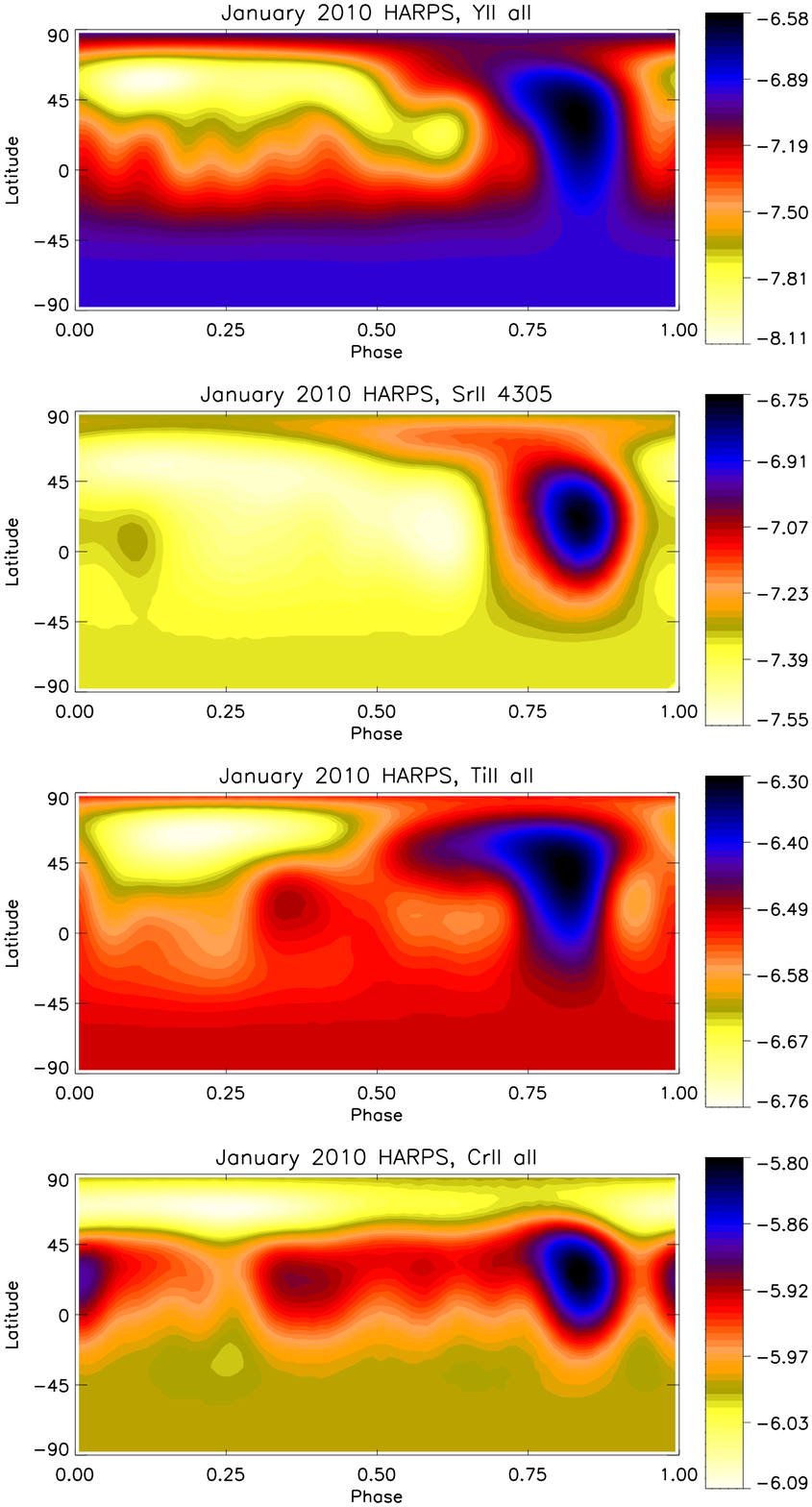}
\caption{Comparison between CORALIE (left) and HARPSpol (right) January 2010 
  maps. The ephemeris is the same as used for CORALIE data in this paper.}
\label{2010maps}
\end{figure}

Figure~\ref{2010maps} shows the result for the 2010 January CORALIE and 
HARPSpol data. The model fits the HARPSpol observations of \ion{Y}{ii} and 
\ion{Cr}{ii} are given in the on-line Figs.~\ref{2010HARPSYspectra} \& 
\ref{2010HARPSCrspectra}, respectively. On the whole the spot configurations in
the maps are virtually identical for the strongly variable lines of \ion{Y}{ii}
and \ion{Sr}{ii}. Also, \ion{Ti}{ii} results are very similar with small 
differences in the lower abundance spot at phases 0.2 to 0.4. The biggest 
differences are seen in the least variable element \ion{Cr}{ii}, where the 
equatorial high abundance spot around the phase 0.0 is not seen in the CORALIE 
data. Otherwise even the \ion{Cr}{ii} maps are very similar. The differences in
the equatorial spot belt around the phases 0.6 to 0.7 can be explained by the 
phase gap in the CORALIE data spanning phases 0.58 to 0.79.

\onlfig{22}{
\begin{figure}
\centering
\includegraphics[width=0.49\textwidth]{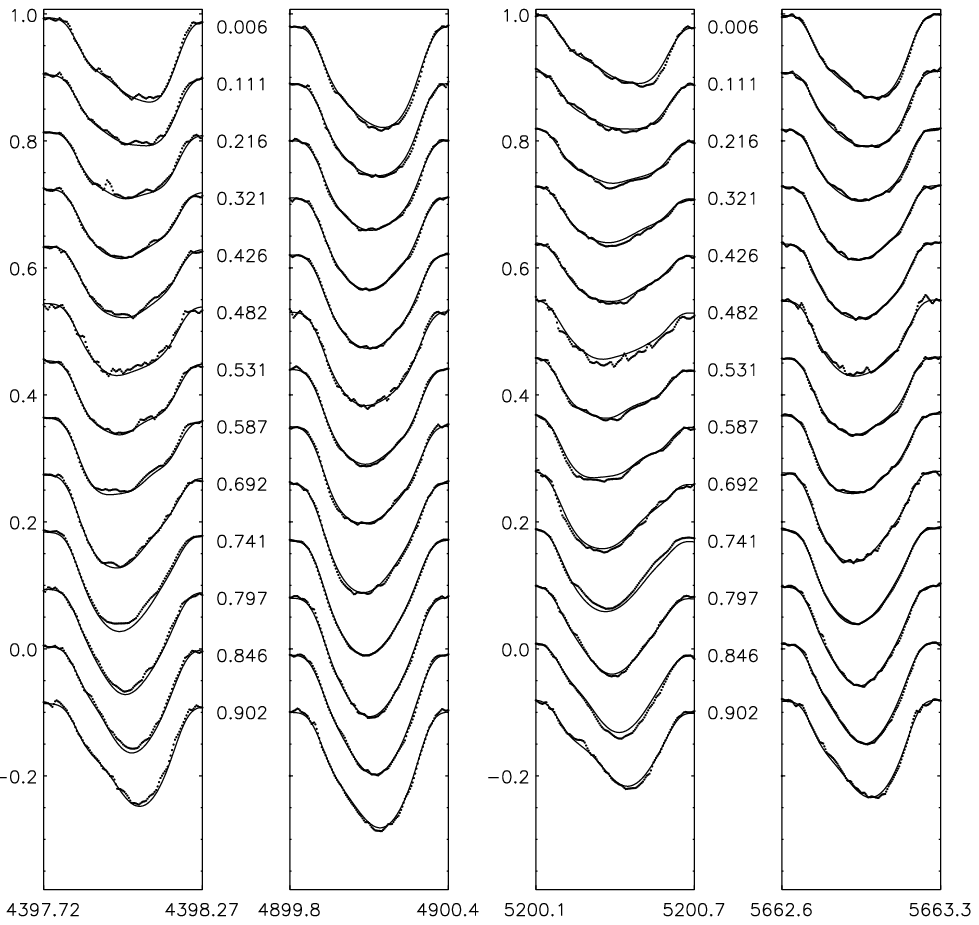}
\caption{Same as on-line Fig. \ref{set1Yspectra} except now for January 2010 
  HARPSpol \ion{Y}{ii} observations.}
\label{2010HARPSYspectra}
\end{figure}}

\onlfig{23}{
\begin{figure}
\centering
\includegraphics[width=0.49\textwidth]{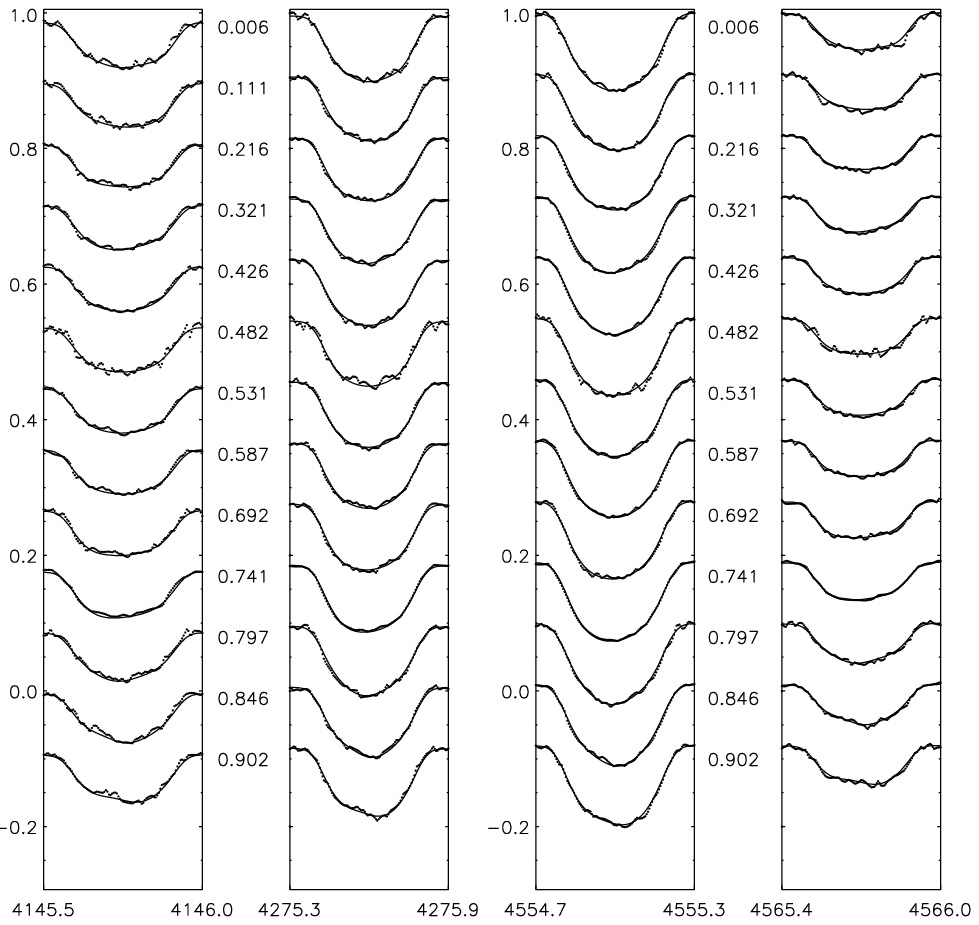}
\caption{Same as on-line Fig. \ref{set1Yspectra} except now for January 2010 
  HARPSpol \ion{Cr}{ii} observations.}
\label{2010HARPSCrspectra}
\end{figure}}

From this comparison it is clear that CORALIE data can be used for Doppler 
imaging of a relatively slowly rotating star, like HD~11753.

\subsection{Impact of phase gaps in the data}
\label{gap}

\begin{figure}
\centering
\includegraphics[width=0.242\textwidth]{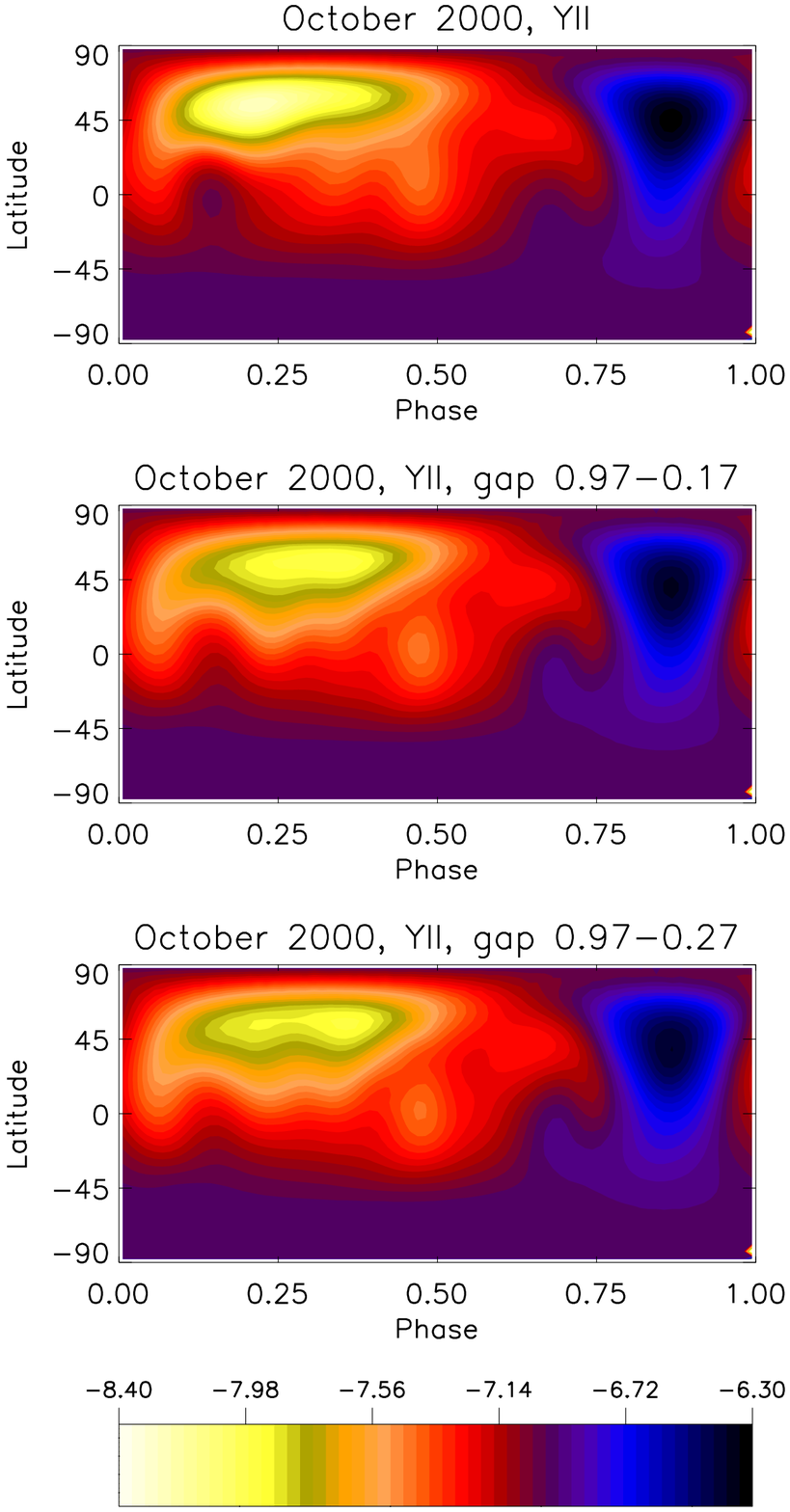}
\includegraphics[width=0.242\textwidth]{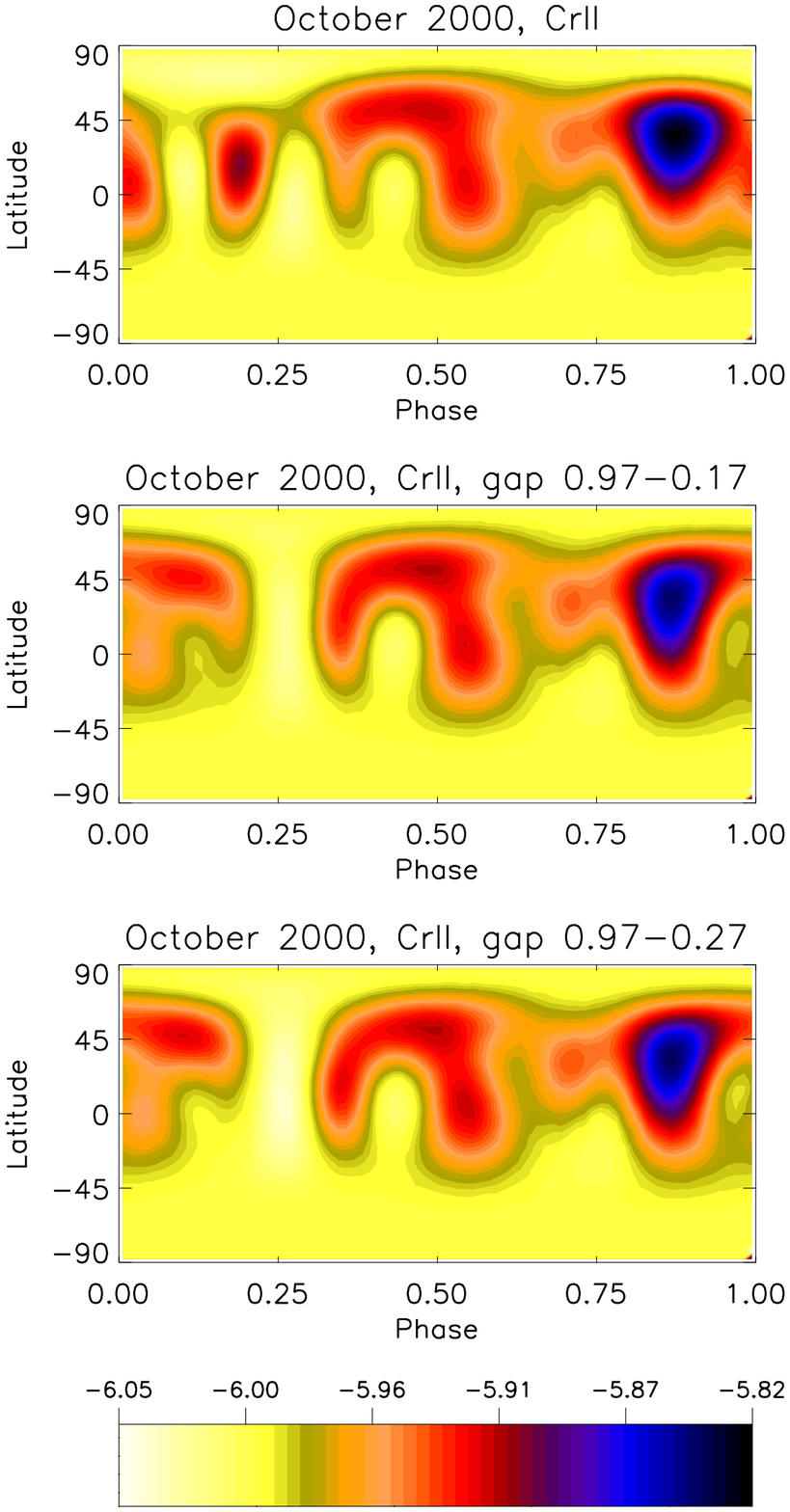}
\caption{Phase gap tests using 2000 October CORALIE data for the \ion{Y}{ii} 
(left) and \ion{Cr}{ii} (right) lines. The top figure is the original map, 
middle one has phases 0.97--0.17, removed and the bottom one phases 0.97--0.27 
removed. }
\label{gaps}
\end{figure}

The dataset from October 2000 has an excellent phase coverage with the largest 
phase gap around 0.1 in phase. Unfortunately, there are larger gaps for the 
other datasets: 0.16 in phase for 2010 January HARPSpol data (0.36--0.52), 
0.20 in phase for December 2000 (phases 0.19--0.39), 0.21 for January 2010 
(phases 0.58-0.79), and 0.30 for August 2009 (phases 0.00-0.30). To study the 
effect of these phase gaps in the maps a test was done using the October 2000 
dataset that removed all the observations between phases 0.00 and 0.20/0.30 to 
simulate the size and location of the largest phase gap (August 2009 data). The
resulting maps for \ion{Y}{ii} and \ion{Cr}{ii} are shown in Fig.~\ref{gaps}. 
The figure shows the original map and maps with the 0.2 and 0.3 phase gaps. 
Tests were carried out using \ion{Y}{ii} and \ion{Cr}{ii} because these are the
elements with the largest and smallest variability, respectively.

Removing the phases 0.00--0.30 affects the exact determination of the shape of 
the spots at this phase range. Still, on the whole the \ion{Y}{ii} map is very 
similar to the original map without the phase gap, the location of the spots 
has not changed, but the high latitude low abundance feature is less prominent.
In the case of \ion{Cr}{ii}, which shows much weaker variability, the recovery 
of spots in the phase gap is seriously affected. The high abundance spot around
the phase 0.18 is basically not recovered, and the exact shape of the high 
abundance spot at the phase 0.05 is also significantly affected.

From this it can be concluded that, especially in the case of lines with weak 
variability, \ion{Ti}{ii} and \ion{Cr}{ii}, one has to be careful when 
interpreting spot features when the phase coverage is not optimal. The recovery
of the surface features in the strongly variable lines is not as affected 
by the phase gaps. Naturally also the exact spot configuration has an effect on
the recovery, because large spots are less affected by phase gaps.

\section{Discussion}

\subsection{Long-term abundance evolution}
\label{long_term}

\begin{figure*}
\centering
\includegraphics[width=0.7\textwidth]{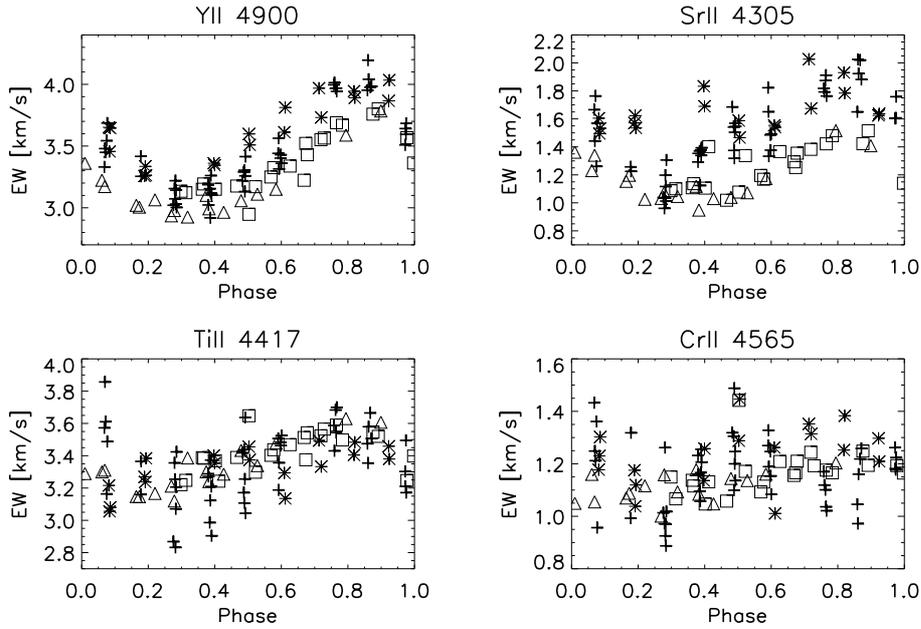}
\caption{Equivalent width measurements from the CORALIE data at different 
epochs. The results for four different elements are shown for October 2000 
(plus), December 2000 (asterisk), August 2009 (square), and January 2010 
(triangle). The ordinates have very different scales.}
\label{EW_all}
\end{figure*}

For investigating long-term changes in the elemental spots in HD~11753 we have 
also measured equivalent widths from all the CORALIE datasets for the lines 
we have used in the inversions. Examples of the results for each element are 
shown in Fig~\ref{EW_all}. 

As already pointed out in Section \ref{sec_period} the equivalent width in 
\ion{Y}{ii} decreases with time. This is seen in all the lines used in the 
inversions. In addition the same effect is clearly seen for \ion{Sr}{ii}. This 
effect is not just present in the 4305.443~{\AA} line used here, but for 
example \ion{Sr}{ii} 4215.519~{\AA} line shows it, too. The diminishing 
abundance is also clearly seen in the \ion{Y}{ii} and \ion{Sr}{ii} maps in 
Figs.~\ref{Ymaps} \& \ref{Srmaps}. On the other hand, no significant long-term
change is seen in the \ion{Ti}{ii} and \ion{Cr}{ii} equivalent widths.

\begin{figure*}
\centering
\includegraphics[width=0.247\textwidth]{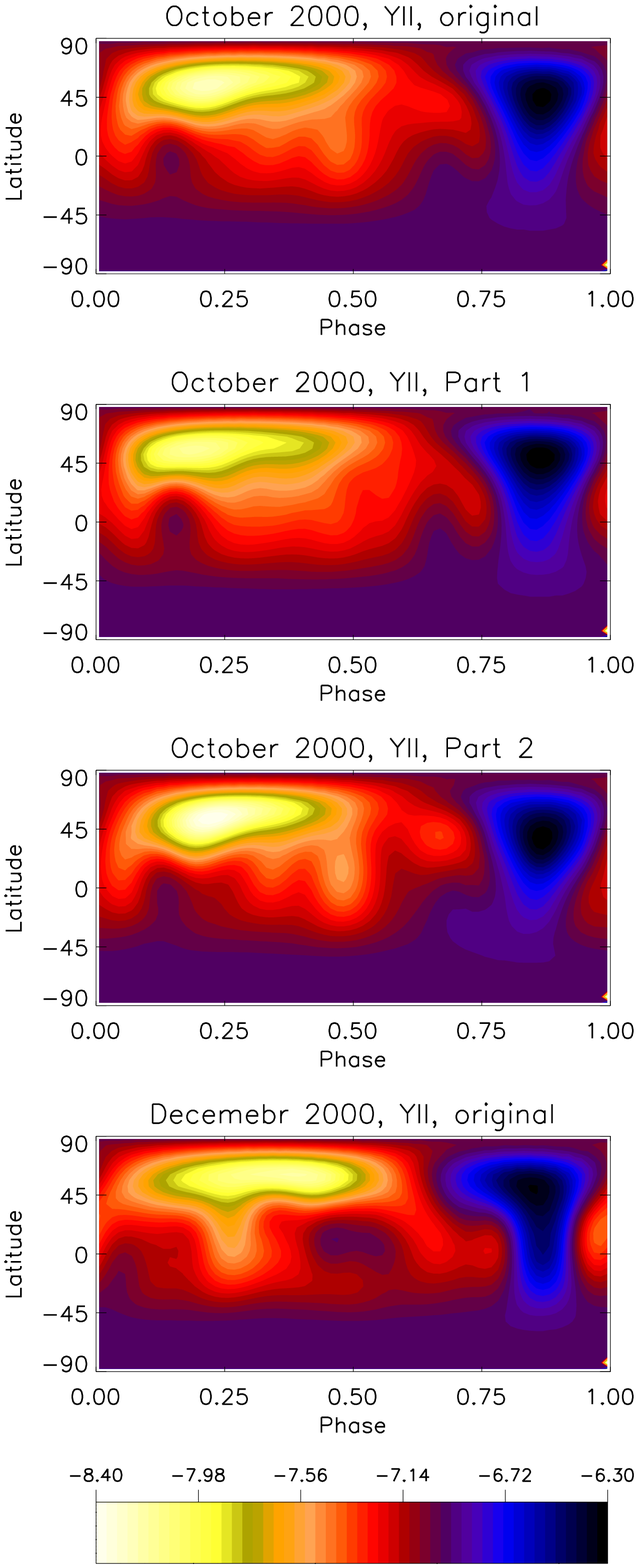}
\includegraphics[width=0.247\textwidth]{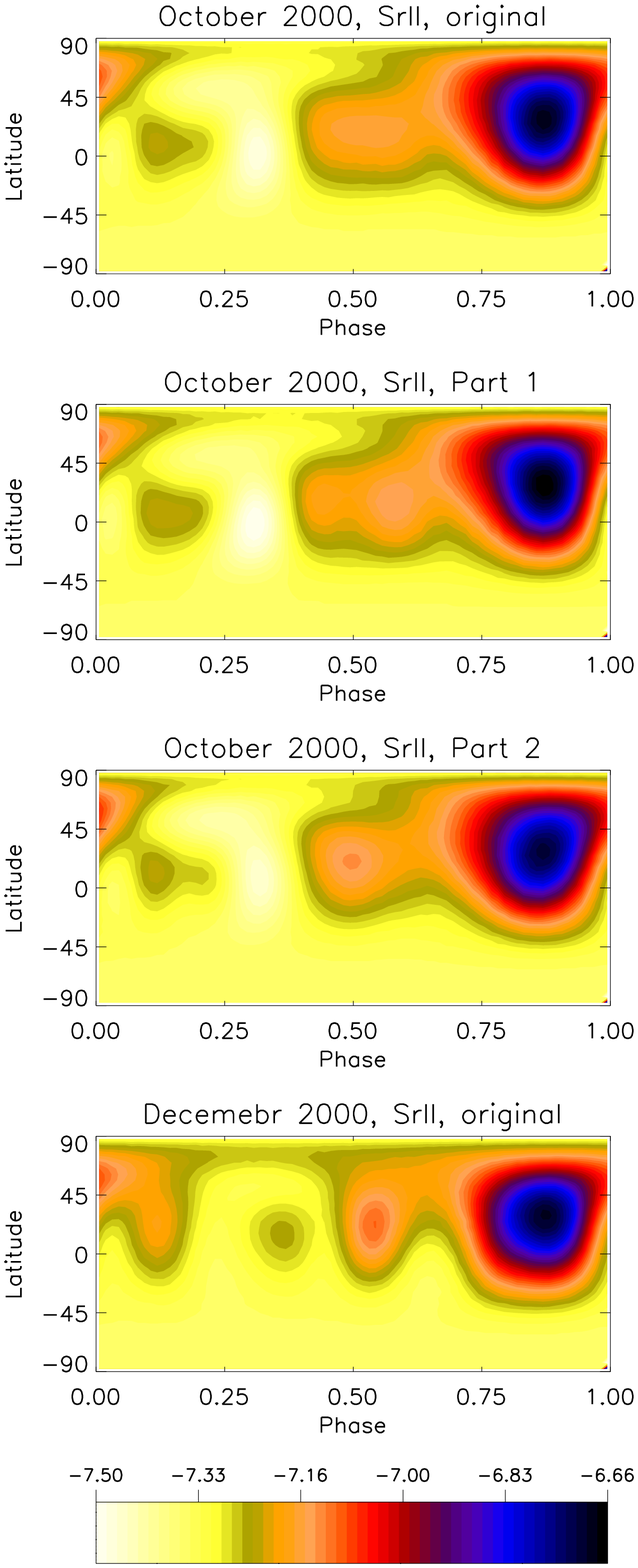}
\includegraphics[width=0.247\textwidth]{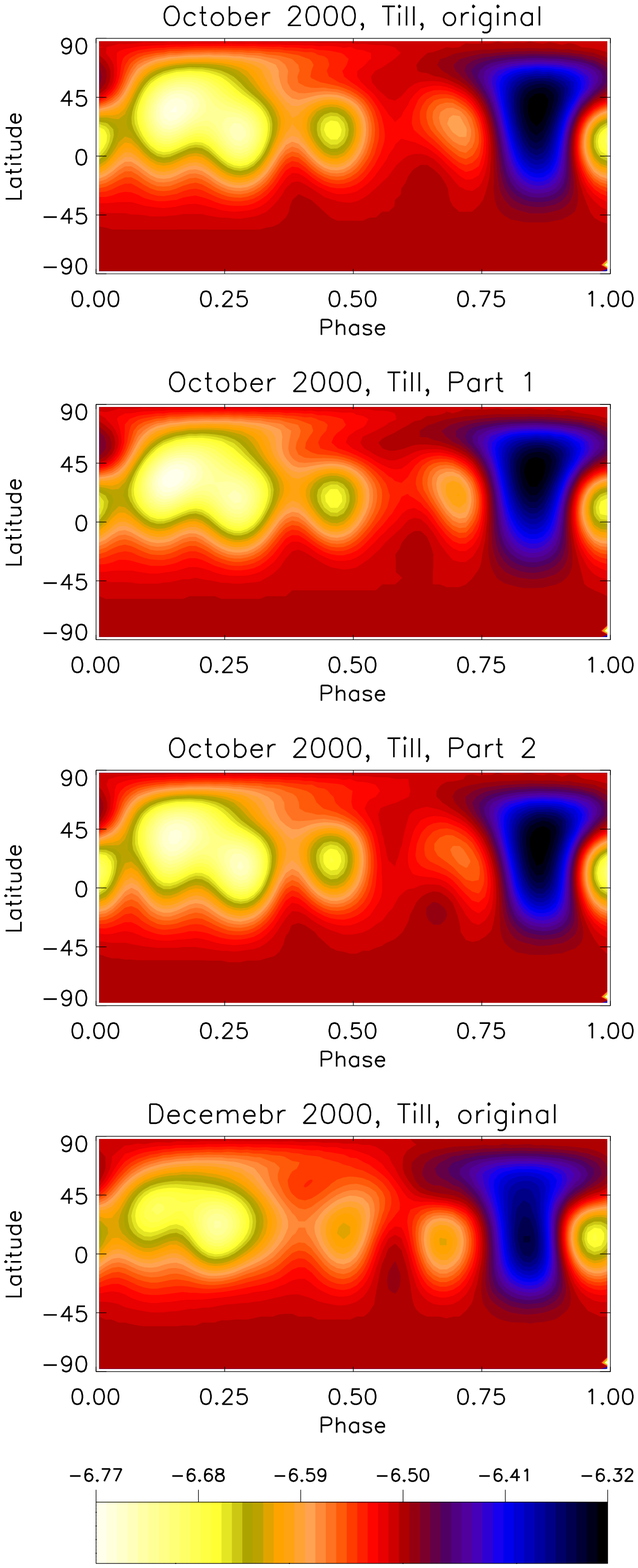}
\includegraphics[width=0.247\textwidth]{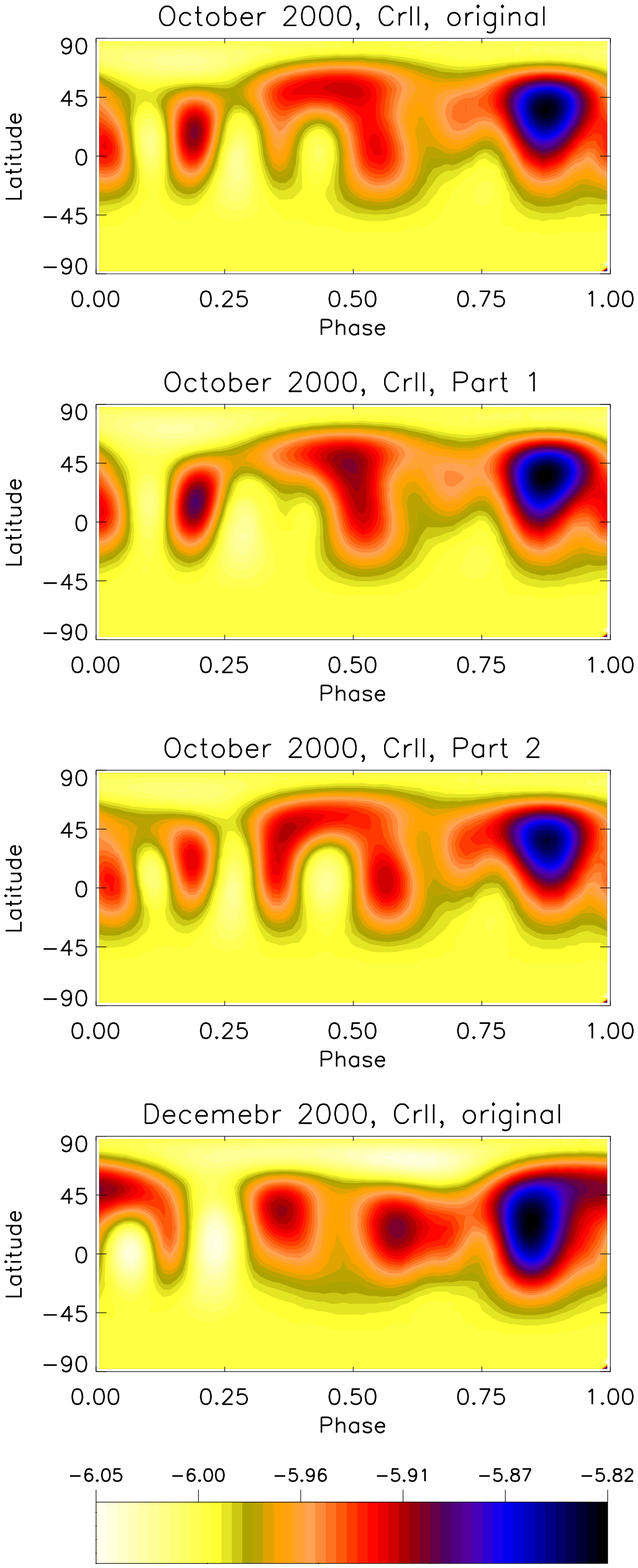}
\caption{Investigating the fast temporal evolution of the elemental spots. 
The figures show original October 2000 map, October 2000 map from half of the 
data, October 2000 map from the other half of the data and the original 
December 2000 for all the elements. The elements are from left to right: 
\ion{Y}{ii}, \ion{Sr}{ii}, \ion{Ti}{ii}, and \ion{Cr}{ii}.}
\label{comp_2000}
\end{figure*}

We want to emphasise that the maps of chemical elements obtained here and the 
conclusions about the evolution of their surface distribution, are not affected
in any significant manner by the possible ambiguity in the rotational period 
(see Fig.\ref{fig:pdg}), since the maps are calculated with observations from 
one relatively short run. In the same manner, the differential rotation 
discussion in Section~\ref{DR_sec} would remain unchanged, since it is based on
a differential analysis of the spot patterns.

\begin{figure*}
\centering
\includegraphics[width=0.99\textwidth]{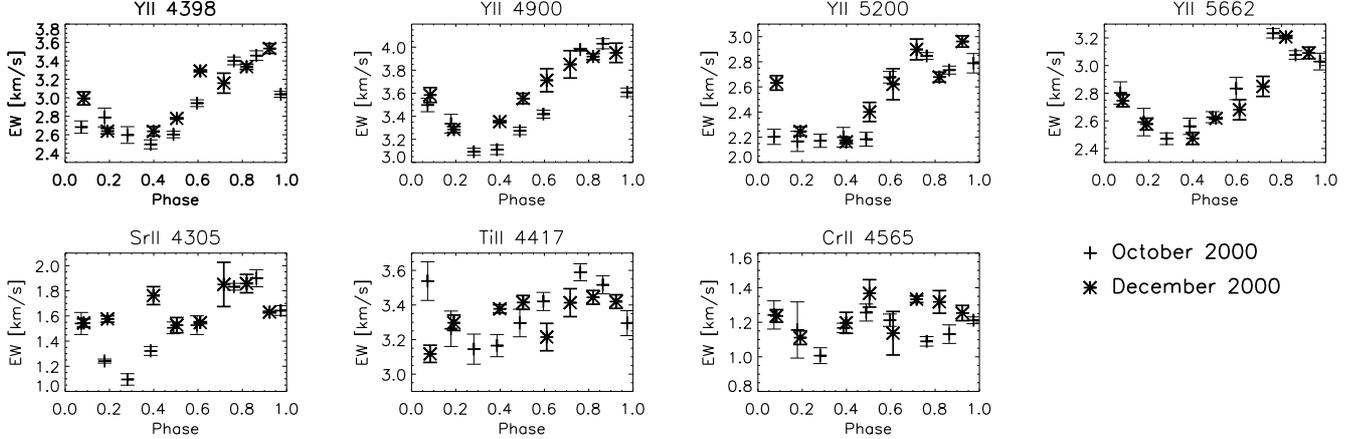}
\caption{Equivalent width measurements from October and December 2000 datasets 
for all the \ion{Y}{ii} and \ion{Sr}{ii} lines used in inversion and examples 
of \ion{Ti}{ii} and \ion{Cr}{ii} measurements. The values are from the first 
stellar rotation in October 2000 (plus) and last rotation in December 2000 
(asterisk). The value is the mean of the observations obtained within one 
night. The error is the standard deviation of the measurements. The ordinates 
have different scales.}
\label{EW_2000}
\end{figure*}

\subsection{Fast chemical spot evolution}

In Paper\,1 fast chemical spot evolution within the 65 days between the October 
2000 and December 2000 maps was reported. Here we look into this in more detail.

To investigate how many random changes one would expect in the maps with the 
current data quality, the October 2000 dataset was divided into two subsets. The
division was done by taking every other phase for part\,1 and the other for 
part\,2, without any overlap in the data. Because our complete dataset for 
October 2000 has excellent phase coverage, and also many observations from 
similar phases (76 phases), this dataset was chosen for the test. The 
effect of the division of 2000 October data into two subsets is effectively 
only lower S/N per phase in the inversion, but with similar phase coverage as 
in the original dataset. 

The maps for all the elements from the original 2000 October dataset, part\,1,  
part\,2, and original 2000 December datasets are shown in Fig~\ref{comp_2000}. 
As can be seen, the original October 2000 map and the maps from parts\,1 and 
2 are virtually identical for all the elements, except for \ion{Cr}{ii} where 
the part\,1 map is missing the equatorial extension of the high abundance
spot around the phase 0.3. Otherwise the parts\,1 and 2 maps of \ion{Cr}{ii} 
are also very similar to the original one.

The abundance scales for all the elements are slightly different for maps 
obtained from part\,1 and part\,2. This implies that the abundance scale in 
general in the maps is on average accurate to $\sim$4\%. In most cases the 
variation in the maximum and minimum abundance in the part\,1 and part\,2 maps 
is 0.2--5.6\% of the full abundance range in the particular map. There are two 
notable differences, though, the maximum \ion{Cr}{ii} abundances in the two maps
show a difference of 7.4\% of the full abundance range, and minimum \ion{Y}{ii} 
abundances show a difference of 12\%.

The December 2000 maps are more different from the October 2000 maps, than the 
maps from the divided datasets are from the original October 2000 maps. This is
true for all the elements, and implies that an evolution of the chemical spots
has occurred during the 65 days between the 2000 October and 2000 December 
datasets.

In \ion{Y}{ii} both the high abundance spot around the phase 0.75 and the lower
abundance spot around the phase 0.25 get less prominent and at the same time
more extended with time. The equatorial extension of the low abundance spot at 
the phase of 0.25 seen in the December 2000 map could be due to missing 
observations at phases 0.19 to 0.39, but it would not explain the extension of 
the low-abundance spot towards later phases, and the wider high abundance spot 
at the higher latitudes.

For \ion{Sr}{ii} the over-all abundance at phases 0.2 to 0.5 is higher in 
December 2000 than it is in October 2000. In addition the high abundance spot 
at phase 0.5 has become more prominent with time. One has to remember, though, 
that the \ion{Sr}{ii} maps are obtained from only one spectral line.

In the case of \ion{Ti}{ii} the low-abundance feature at the phase 0.5 does
not have as low an abundance in December 2000 as it has in October 2000, and 
the low abundance feature around the phase 0.7 is more prominent in December 
2000. \ion{Cr}{ii}, on the other hand, exhibits an extended low-abundance 
high-latitude feature around the phase 0.5 in the December 2000 map. This 
feature is not present in the October 2000 \ion{Cr}{ii} map. One should keep in
mind that the spectral variability is relatively weak in \ion{Ti}{ii} and 
\ion{Cr}{ii}.

We also investigate whether the effects can be seen in the equivalent width
measurements. This is more demanding, because the behaviour of equivalent width 
does not directly reflect the element distribution, but is an integral value and
can appear almost constant even if the distribution of the element is slightly 
changing on the surface. Figure \ref{EW_2000} shows the equivalent width 
measurements for the 2000 October and 2000 December datasets in all the 
\ion{Y}{ii} and \ion{Sr}{ii} lines used in the inversions, and also examples 
for \ion{Ti}{ii} and \ion{Cr}{ii} lines. The 2000 October measurements are from
the first full stellar rotation of that dataset. Each datapoint is the mean of 
all the observations obtained within one night, and the standard deviation of 
the measurements provides the error. The 2000 December measurements are treated
similarly, but now the last full stellar rotation is used.

In general the equivalent width measurements for October 2000 and December 2000 
are very similar, but there are some differences. The equivalent widths for 
\ion{Y}{ii} 4900.120~{\AA} in December 2000 are clearly larger in phases 
0.4 to 0.6 than they are in October 2000 at the same phases. Similar trends are 
also seen in \ion{Y}{ii} 4398.013~{\AA} and 5200.405~{\AA}, but they are less 
prominent. Equivalent widths for \ion{Sr}{ii} 4305.443~{\AA}, on the other 
hand, show very clear differences at phases 0.2 to 0.5, with higher values in 
December 2000. This behaviour is also seen in \ion{Sr}{ii} 4215.519~{\AA}, 
which was checked for consistency, and is in line with what is seen in the 
\ion{Sr}{ii} maps for the year 2000 (Fig.~\ref{comp_2000}). Both \ion{Ti}{ii} 
4417.714~{\AA} and \ion{Cr}{ii} 4565.739~{\AA} lines show much less variability
than the lines of the other elements, but the tendency for the measurements at 
phases 0.4--0.6 to have higher values in December 2000 than in October 2000 is 
still seen in \ion{Ti}{ii} 4417.714~{\AA}.

\begin{figure*}
\centering
\includegraphics[width=0.247\textwidth]{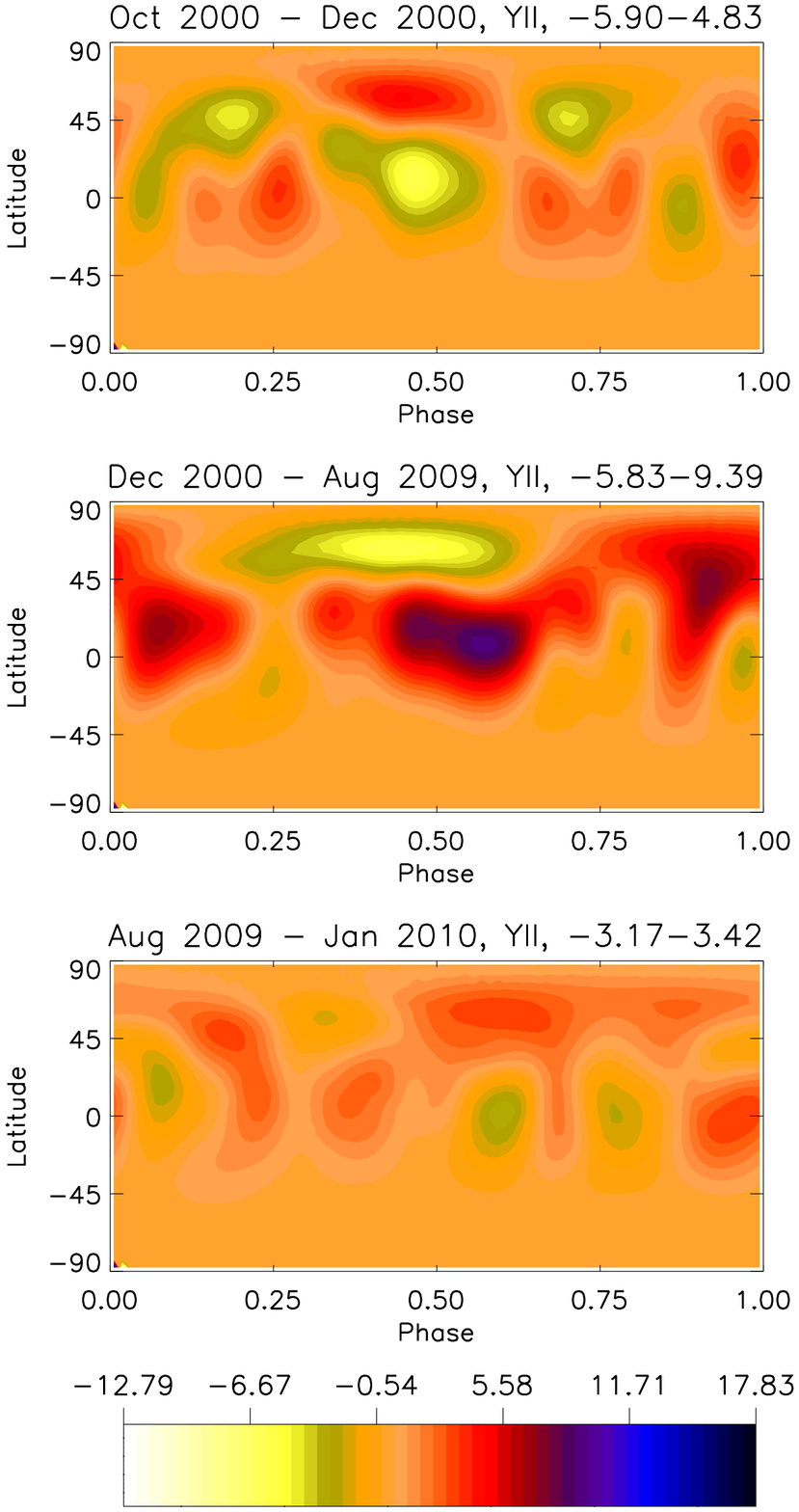}
\includegraphics[width=0.247\textwidth]{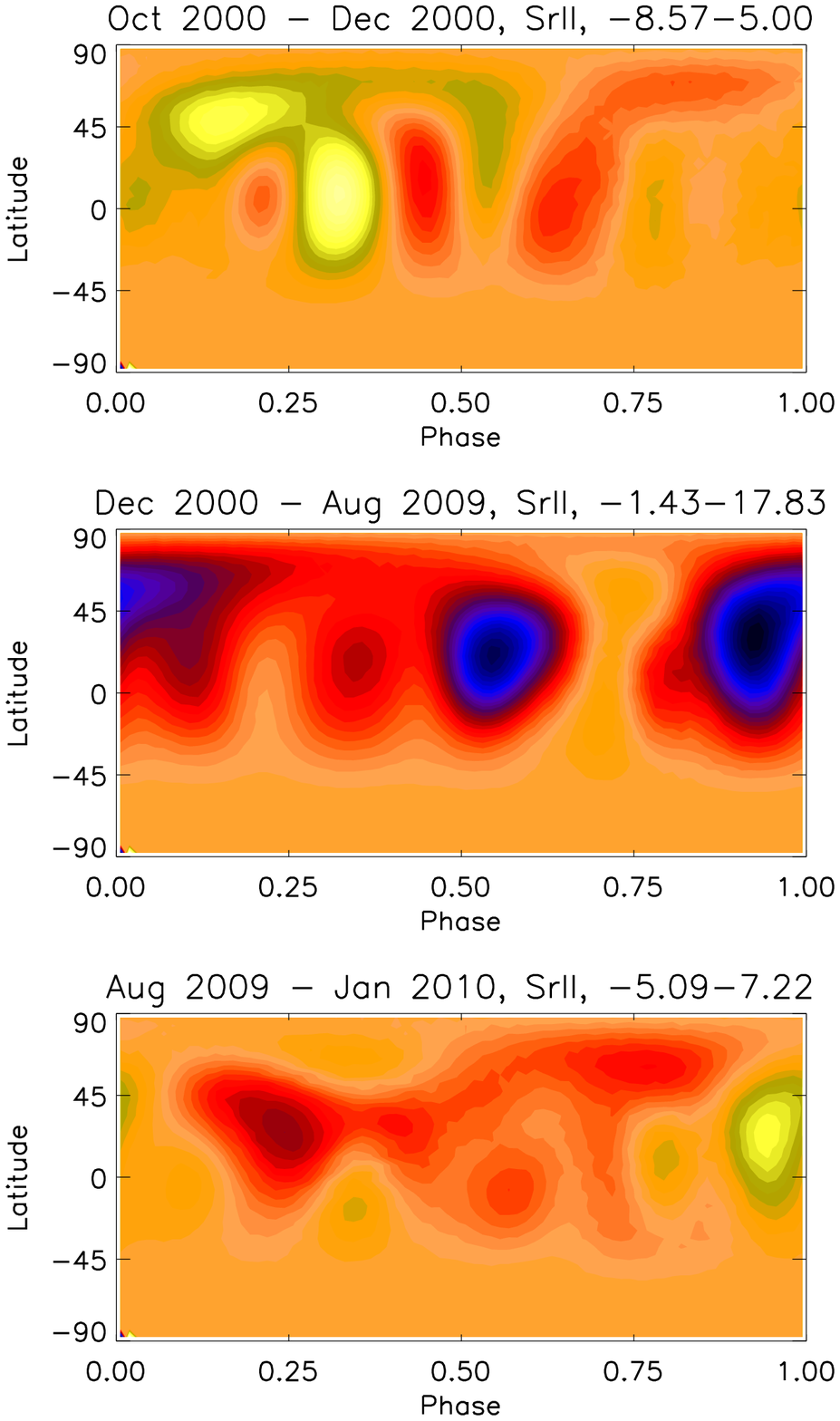}
\includegraphics[width=0.247\textwidth]{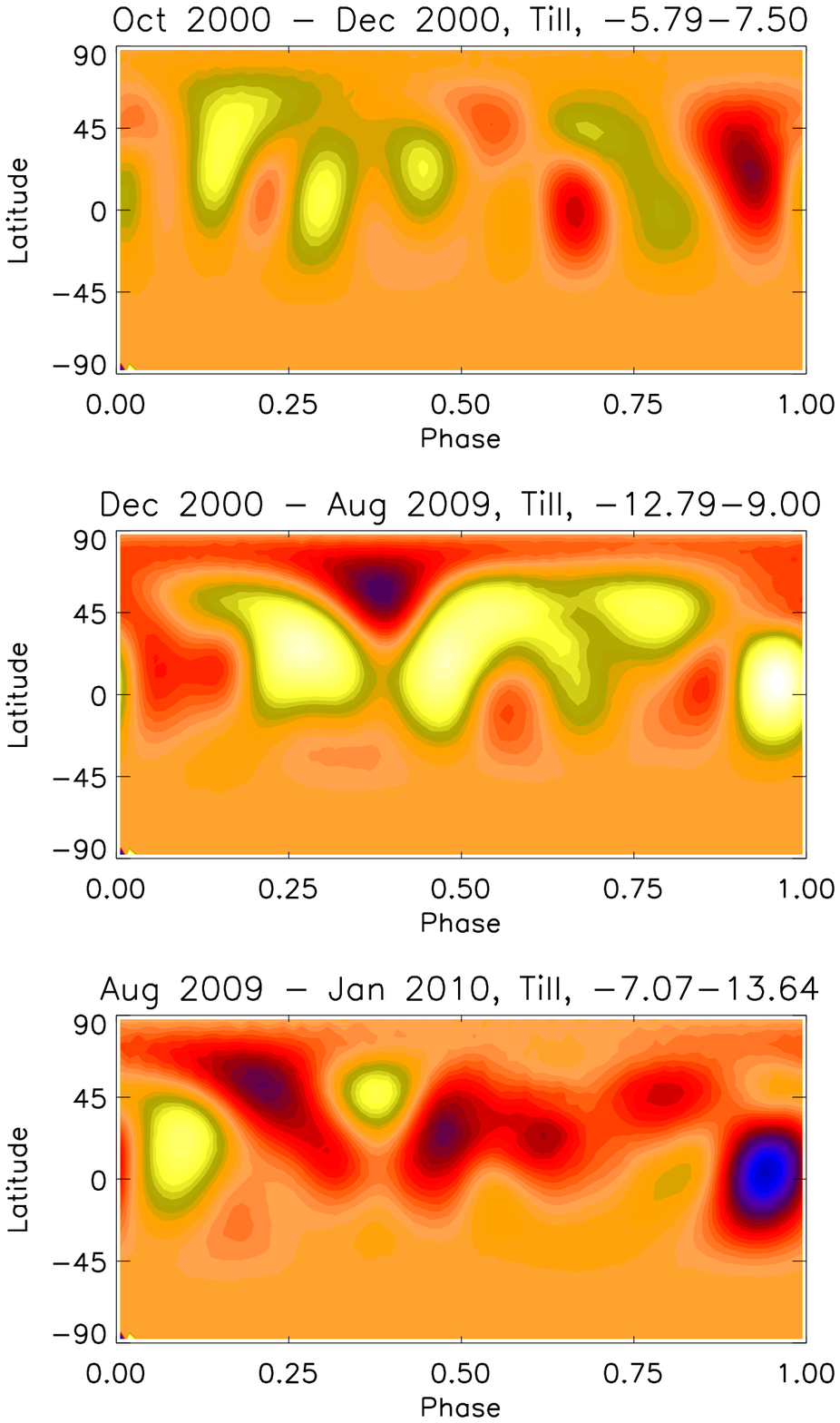}
\includegraphics[width=0.247\textwidth]{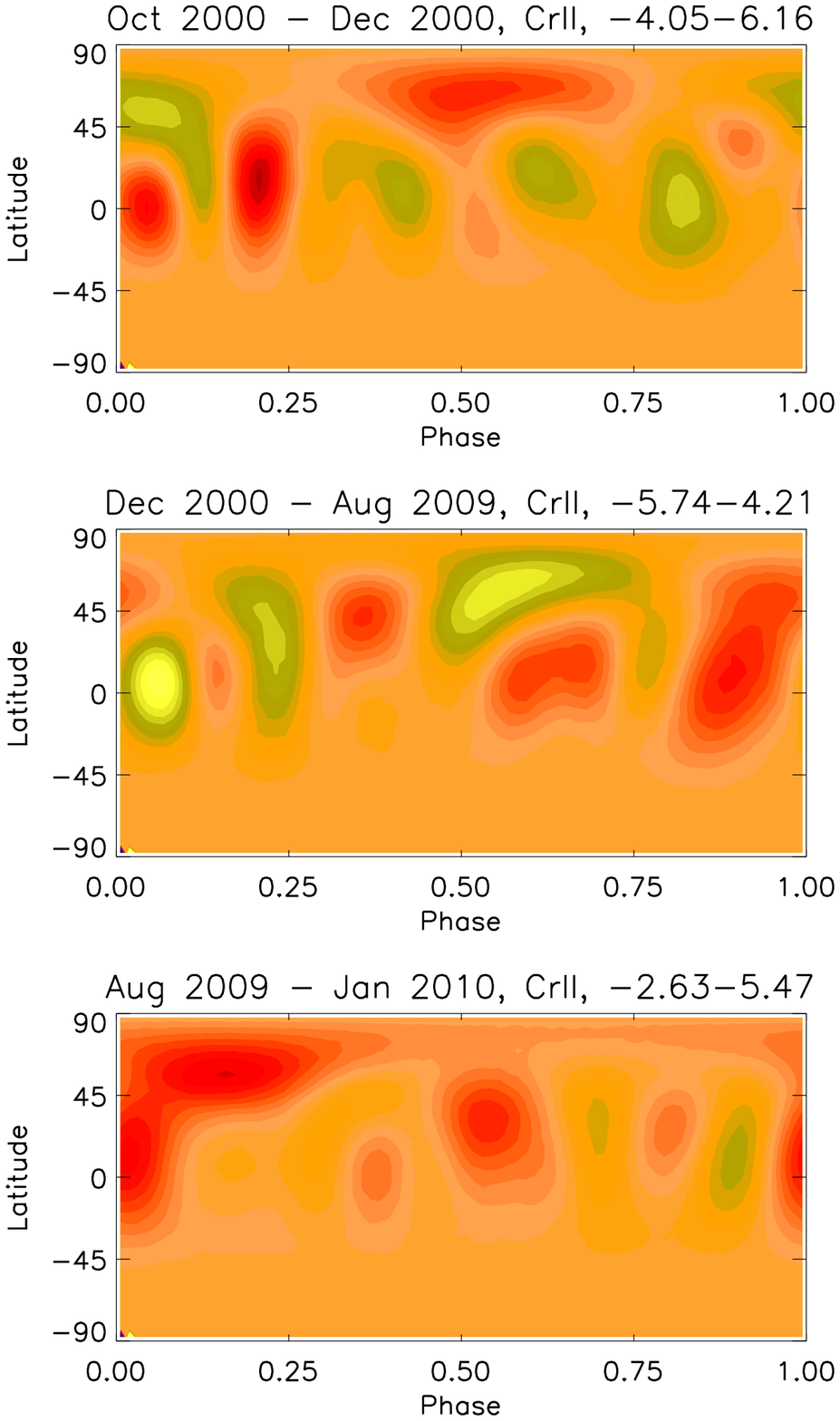}
\caption{The significance of the changes between two consecutive epochs. The 
elements are from left to right: \ion{Y}{ii}, \ion{Sr}{ii}, \ion{Ti}{ii}, and 
\ion{Cr}{ii}. The figure gives standard deviation of the change for the 
following epochs: October 2000 - December 2000, December 2000 - August 2009, and
August 2009 - January 2010, from top to bottom. The original difference maps 
have been divided by the standard deviation of the difference for the October 
2000 part 1 -- part2 map, to present the significance of the changes. All the 
significance maps have the same scale, and the scale for each map is given in 
the title of the individual plot. Colour-coding is such that the dark colour 
indicates higher abundance in the first map from which the second map is
subtracted; similarly, bright colour indicates lower abundance in the first
map.}
\label{diff}
\end{figure*}

Finally, the significance of the detected changes was evaluated. First, 
difference maps between consecutive epochs for all the elements were 
calculated, e.g., subtracting the abundances of December 2000 map from the 
abundances of October 2000 map for \ion{Y}{ii}. The amount of changes seen in
 the difference maps of the two October 2000 submaps (parts\,1 and 2 from Fig. 
\ref{comp_2000}) is taken to be the detection limit of the variation. In the 
significance maps presented in Fig.~\ref{diff}, the difference maps are divided 
by the standard deviation of the difference part1 - part2 of that element to 
show the standard deviation of the change. The plot gives October 2000 - 
December 2000, December 2000 - August 2009, and August 2009 - January 2010, from
top to bottom, and \ion{Y}{ii}, \ion{Sr}{ii}, \ion{Ti}{ii}, and \ion{Cr}{ii} 
from left to right.

\begin{table}
\caption{Standard deviation of the difference maps.}
\label{map_stdv}
\centering 
\begin{tabular}{ r c c c c}
\hline\hline            
Difference map & \ion{Y}{ii} & \ion{Sr}{ii} & \ion{Ti}{ii} & \ion{Cr}{ii} \\ 
\hline    
Oct 2000: part1 -- part2   & 0.077 & 0.023 & 0.014 & 0.019 \\
Oct 2000 -- Dec 2000 & 0.132 & 0.044 & 0.029 & 0.031 \\
Dec 2000 -- Aug 2009 & 0.215 & 0.091 & 0.049 & 0.034 \\
Aug 2009 -- Jan 2010 & 0.087 & 0.045 & 0.042 & 0.026 \\
\hline                       
\end{tabular}
\end{table}

In addition, the standard deviation of each difference map is calculated and 
results given in Table~\ref{map_stdv}. The changes seen in the earlier 
analysis of the maps and equivalent width measurements are confirmed by the 
difference maps. In most cases the standard deviation of the difference map is 
twice, or more, than of the deviation part1 - part2 for that element. 

Subtracting the December 2000 map from the October 2000 one in \ion{Y}{ii} 
still indicates that the high-latitude, lower abundance spot at phases 0.0--0.4 
extends more towards phase 0.5 in the December 2000 map, and that at the 
same time the lower abundance spot has become less prominent. The biggest 
differences in the \ion{Y}{ii} abundance are seen in the equatorial region 
around phase 0.5. This is caused by a high abundance spot at this location 
in the December 2000 map, that is not present in the October 2000 map, nor in 
the August 2009 map. Also, the virtual disappearance of the lower abundance 
spot at high latitudes around phases 0.2 to 0.4 between December 2000 and 
August 2009 is also confirmed. This change cannot be completely explained by 
the missing phases 0.0 to 0.3 in the August 2009 map, because the difference is 
clearly seen also at phases 0.3 to 0.6. In addition, it is clear that in many 
regions the abundance in the August 2009 map is lower than in the December 2000
map. On the other hand, there are basically no significant changes between 
August 2009 and January 2010 in \ion{Y}{ii}.

In the \ion{Sr}{ii} difference maps, the changes between October 2000 and 
December 2000 are confirmed, and the higher abundance in December 2000 around 
the phase 0.3 is seen well. The overall change in the abundance between 
December 2000 and August 2009 is clear, and especially the high abundance spot 
around the phase 0.75 becomes less extended. Again the changes between August 
2009 and January 2010 are small, with the higher abundance around phases 
0.1 to 0.3 in the August 2009 map being the main difference. One has to 
remember, though, that in August 2009 map the observations are missing from 
phases 0.0 to 0.3, and the difference in the abundances could be caused by the 
phase gap.

In the \ion{Ti}{ii} difference maps the largest changes are seen in the 
abundance of the main low abundance spot around the phase 0.9. Changes in the 
exact spot configurations in other locations are also seen, and it seems that 
the August 2009 map is different from the others. This is verified by 
subtracting January 2010 map from the December 2000 map; the differences are 
clearly smaller than between December 2000 and August 2009. \ion{Cr}{ii} 
difference maps imply very small changes in the exact spot configurations at a 
wide range of locations. 

\subsection{Investigating possible surface differential rotation}
\label{DR_sec}

We applied the \emph{spot-centre tracking }technique for the detection of 
stellar velocity fields to the time series of Doppler maps. The method will be 
tested and explained in detail in Flores Soriano et al. (in preparation). We 
describe here the basics of the process and the steps taken.

To reduce the effects of spot evolution and artefacts, only spots that can be 
unambiguously identified at the five different epochs (maps from CORALIE 
data and also from HARPSpol data presented in Section \ref{comp}) are used. This
limits our analysis to the most prominent overabundance features. The border of
the spot is chosen as the abundance that maximises its size without been 
affected by neighbour structures or potential artefacts. In this case, we chose 
$-5.89$ for Cr, $-7.1$ for Sr, $-6.44$ for Ti, and $-6.8$ for Y. When spurious 
spots are also isolated, they are removed and not considered in the analysis. 
Since all spots are at a similar latitude, it is not possible to unambiguously 
detect a latitudinal dependence of the rotation period by measuring the 
motion of the spots as a whole. Nevertheless, their large extent means it 
remains viable to calculate it by measuring the rotation period in different 
latitude bands. For doing this the maps were divided into equal-latitude 
strips, each one with a width of 4.5$^\circ$, equal to the pixel size in the 
maps. For the region of the spot enclosed in each band, the coordinates of its 
centre are calculated as an ``\emph{abundance centre}'' (in analogy to the 
centre of mass) where areas with an abundance lower than the selected limit 
make no contribution. Latitudinal rotation rates are calculated by fitting the 
evolution of the longitude as a function of time to a line.

\begin{figure}
\centering
\includegraphics[width=0.35\textwidth]{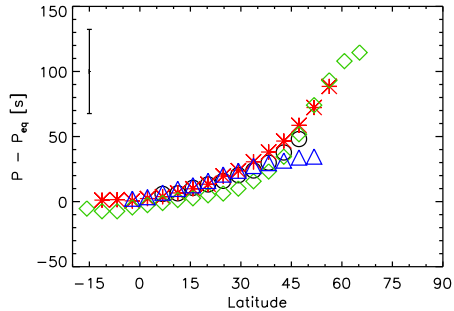}
\caption{Relative rotation periods at different latitudes measured from 
different chemical elements: \ion{Y}{ii} measurements are marked with (blue) 
triangles, \ion{Sr}{ii} with (red) asterisk, \ion{Ti}{ii} with (green) 
diamonds, and \ion{Cr}{ii} with (black) circles. The relative rotation periods 
at different latitudes obtained from different elements are remarkably similar.
The average errorbar of the measurements is given in the upper left corner.}
\label{rotation}
\end{figure}

\begin{figure}
\centering
\includegraphics[width=0.35\textwidth]{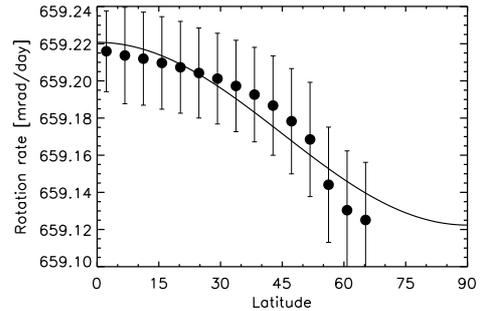}
\caption{Surface rotation of HD~11753. Latitudinal rotation rates averaged from
measurements of different elements at the same latitude. A $\sin^{2}$-law has 
been fitted to the measurements. The results imply very weak solar-type surface
differential rotation. }
\label{DR}
\end{figure}

According to our results, the rotation period is minimum near the equator and 
gradually increases until latitude 65$^\circ$, where it reaches a value that is 
$\sim$115 seconds higher. As can be seen in Fig.~\ref{rotation}, the rotation 
periods relative to the mean rotation period obtained from all the maps 
(\ion{Y}{ii}, \ion{Sr}{ii}, \ion{Ti}{ii}, and \ion{Cr}{ii}) are in excellent 
agreement with each other. The relative rotation periods from different 
elements at the same latitude are averaged, and the resulting profile fitted 
with the usual differential rotation law:
\[
\Omega\left(\theta\right)=\Omega_{\textrm{eq}}-\Delta\Omega\sin^{2}\theta
\]
where $\theta$ is the latitude, $\Omega_{\textrm{eq}}$ the equatorial
rotation rate, and $\Delta\Omega=\Omega_{\textrm{eq}}-\Omega_{\textrm{pole}}$. The 
results are also shown in Fig.~\ref{DR}. We obtain a surface shear of 
$\Delta\Omega=0.098\pm0.023$\,mrad/day, equivalent to a lap time (time the 
equator needs to lap the pole) of approximately 175 years. The derived 
equatorial rotation period is $9.53123 \pm 0.00012$ days, i.e., 40 seconds 
longer than found with equivalent widths (see Section \ref{sec_period}). 
Evidence of differential rotation in late B-type star has been found before 
only in one target, the B9 star HD~174648 (Degroote et al. \cite{Degroote}). 

\section{Conclusions}

From the investigation of high-resolution spectra of HD~11753 in four different
epochs, the following conclusions can be drawn:

\begin{itemize}
\item We determined the binary orbit of HD~11753. The radial velocity
measurements can be explained by a wide and eccentric orbit with orbital period 
of 1126 days.
\item We fine-tuned the rotation period determination of HD~11753 using 
data spanning almost ten years.
\item HD~11753 clearly exhibits inhomogeneous distribution of \ion{Y}{ii},
\ion{Sr}{ii}, \ion{Ti}{ii}, and \ion{Cr}{ii}, with the main high abundance 
feature occurring at the same phase in all the elements.
\item The most prominent features in the maps remain at similar location for 
the duration of the observations (10 years), but the exact shapes and 
abundances change.
\item Both fast and secular evolution is seen in the spot configurations, with 
the spot configurations changing even on monthly time scales, and the 
mean abundance of \ion{Y}{ii} and \ion{Sr}{ii} changing on yearly time scales.
\item Some indications of very weak surface differential rotation is seen 
using the spot-centre tracking technique.
\end{itemize}

\begin{acknowledgements}
HK acknowledges the support from the European Commission under the Marie Curie 
IEF Programme in FP7. SH and JFG acknowledge the support by the Deutsche 
Forschungsgemeinschaft (Hu532/17-1). The authors wish to thank Gaspare Lo Curto
from ESO/Garching for his help with the ESO HARPS pipeline when reducing the 
HARPSpol data obtained from the archive. The authors also thank the 
referee, Dr. John Landstreet, for his comments that helped to improve this 
paper.

\end{acknowledgements}

\end{document}